\documentclass[a4paper,UKenglish,cleveref, autoref]{lipics-v2021}

\usepackage{amsmath, amssymb} 
\usepackage{stmaryrd} 
\usepackage{wasysym}
\usepackage{amsthm} 
\usepackage[bibliography=common]{apxproof}
\newtheoremrep{lemma_}[theorem]{Lemma}
\newtheoremrep{theorem_}[theorem]{Theorem}
\newtheoremrep{observation_}[theorem]{Observation}
\newtheoremrep{proposition_}[theorem]{Proposition}

\usepackage{tikz}
\usepackage{mdframed}
\usepackage{xspace}
\usepackage[linesnumbered,lined,boxed,commentsnumbered,noend]{algorithm2e}

\newcommand{\maxcommonpb}{\textsc{Maximum Common Labeled Contraction}\xspace}
\newcommand{\contractibilitypb}{\textsc{Labeled Contractibility}\xspace}
\newcommand{\T}{\mathcal{T}}

\newcommand{\X}{\mathcal{X}}
\newcommand{\Y}{\mathcal{Y}}

\newcommand{\W}{\mathcal{W}}
\newcommand{\cw}{\textrm{cw}}
\newcommand{\sbinpacking}{\textsc{Unary Perfect Bin Packing}\xspace}

\title{The Parameterized Landscape of Labeled Graph Contractions} 

\author{Manuel Lafond}{Department of Computer Science, University of Sherbrooke, Québec, Canada}{manuel.lafond@usherbrooke.ca}{0000-0002-5305-7372}{}

\author{Bertrand Marchand}{Department of Computer Science, University of Sherbrooke, Québec, Canada}{bertrand.marchand@usherbrooke.ca}{0000-0001-8060-6640}{}

\authorrunning{M.\ Lafond and B.\ Marchand} 

\Copyright{Manuel Lafond and Bertrand Marchand} 



\category{} 

\relatedversion{} 

\acknowledgements{}

\nolinenumbers 

\EventEditors{Pat Morin and Eunjin Oh}
\EventNoEds{2}
\EventLongTitle{19th International Symposium on Algorithms and Data Structures (WADS 2025)}
\EventShortTitle{WADS 2025}
\EventAcronym{WADS}
\EventYear{2025}
\EventDate{August 11--15, 2025}
\EventLocation{York University, Toronto, Canada}
\EventLogo{}
\SeriesVolume{349}
\ArticleNo{40}

\begin{document}

\maketitle

\begin{abstract}
    In this work, we study the problem of computing a 
    \emph{maximum common contraction} of two vertex-labeled graphs,
    i.e.\ how to make them identical by contracting as little edges as possible in the two graphs.
    We study the problem from a parameterized complexity point of view,
    using parameters such as the maximum degree, the
    degeneracy, the clique-width or treewidth of the input graphs as well as
    the number of allowed contractions.
    We put this complexity in perspective with that of the \emph{labeled contractibility
    problem}, i.e\ determining whether a labeled graph is a contraction of another.
    Surprisingly, our results indicate very little difference between these
    problems in terms of parameterized complexity status. We only prove
    their status to differ when parameterizing by both the degeneracy
    and the number of allowed contractions, showing W[1]-hardness
    of the maximum common contraction problem in this case, whereas
    the contractibility problem is FPT.
%
%
%
\end{abstract}

\section{Introduction}

Graphs are used as an abstract model in a wide
variety of applicative fields. For instance, in bioinformatics,
they can represent structured RNAs~\cite{rinaudo2012tree,zhao2008rapid,marchand2022tree,boury2023automatic},
evolutionary histories~\cite{huson2010phylogenetic}, or interaction
networks~\cite{lacroix2008introduction}. In several applications, 
 a common
computational task is the quantitative \emph{comparison} of
graphs, in order to underline a common structure.
For example, the comparison of \emph{phylogenetic networks} (graphs
representing evolutionary histories) has a rich line of work
in evolutionary
bioinformatics~\cite{cardona2024comparison,marchand2024finding,huson2010phylogenetic,cardona2008metrics,cardona2008metricsb,gambette2017rearrangement,landry2022defining}. 
Likewise, the \emph{network alignment problem} has been
formulated to compare biological interaction networks~\cite{denielou2009multiple,faisal2015post,kalaev2008fast}.
As for the comparison of molecules, it has motivated work
on the \emph{maximum common subgraph} problem~\cite{abu2017complexity,akutsu2020improved,ehrlich2011maximum}.
The 
\emph{graph edit distance}~\cite{gao2010survey} was also studied for this purpose,
especially in a context of image processing and machine learning~\cite{serratosa2021redefining,zeng2009comparing}.

The philosophy of the graph edit distance is to compare two input graphs
by counting the number of ``edit operations'' required
to transform one graph into the other. 
This edition process can amount to ``reducing'' the first
graph to a ``maximum common reduced graph'' before applying ``augmenting''
operations to get the second graph. For instance, 
the computation of the graph edit distance, given its use of the vertex
deletions and insertions, amounts to
the computation
of a maximum common subgraph under a special cost function~\cite{bunke1997relation}. 
Following this philosophy, the \emph{edge contraction} operation
has been used to define one of the most widely-used distance on evolutionary trees,
called the Robinson-Foulds distance~\cite{robinson1981comparison},
whose computation finds a \emph{maximum common contraction}
of the two trees. Recently,
this approach was generalized to the comparison of phylogenetic
networks~\cite{marchand2024finding}, which are directed
graphs representing evolution. However, as far as the authors
know, the problem of computing a maximum common contraction between
two undirected graphs has received little to no attention from an algorithmic perspective.
There are several deterring factors that may explain this.  First, if the
graphs are unlabeled, then determining whether no contractions are required is
equivalent to checking whether they are isomorphic, a notoriously difficult
problem to tackle.
To make matters worse, even if the input graphs are two (unlabeled) trees,
determining whether one is a contraction of another is
NP-hard~\cite{matouvsek1992complexity}. On the other hand, this hardness does
not apply to \emph{labeled} trees, where each vertex has a unique
identifier, since polynomial-time algorithms exist (even just
\emph{leaf-labeled}
trees as in~\cite{robinson1981comparison}).

\begin{figure}
    \centering
    \includegraphics[width=\textwidth]{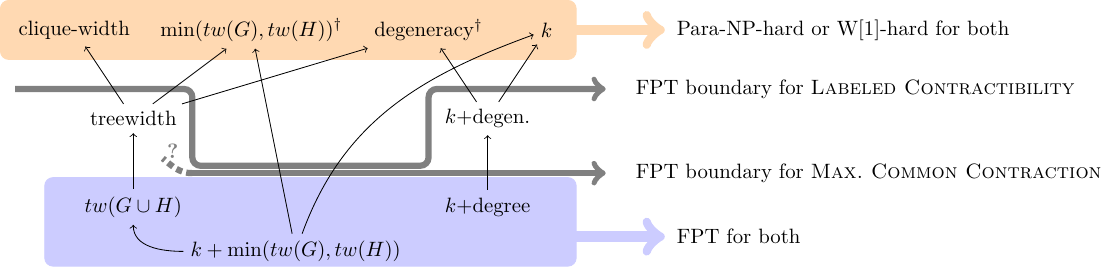}
    \caption{Illustration of our results on the compared parameterized complexity
    of \textsc{Labeled Contractibility} and \textsc{Maximum Labeled Common
    Contraction}. Results marked with a $^\dagger$ are derived from
    the litterature~\cite{marchand2024finding,brouwer1987contractibility}. 
    The clique-width, degeneracy and treewidth are to be understood
    as the maximum value of these parameters on the two input graphs.
    An arrow from (1) to (2) indicates that bounding (1) implies
    bounding (2).}
    \label{fig_results}
\end{figure}

In this article, we study the \maxcommonpb, i.e.\ 
the problem of computing a maximum-size common contraction of two fully-labeled graphs,
from the perspective
of \emph{parameterized complexity}.
The input graphs are uniquely labeled (two distinct vertices of a graph have
distinct labels), but each graph may have labels not present in the other.
We use structural
parameterizations, such as the treewidth or clique-width of the input graphs,
but also the maximum degree, degeneracy or number of contractions. 
We also study the \emph{contractibility} problem (given two labeled graphs,
is one a contraction of the other ?) and compare the parameterized complexity aspects
of both problems. Our results are summarized on Figure~\ref{fig_results},
and outline little difference in complexity between both problems.
We fully establish the location of the barrier between FPT and
W[1]-hardness/para-NP-hardness for our chosen parameters --- with the notable
exception of the maximum common contraction problem when both graphs have
bounded treewidth, which remains open.  We see that the two problems behave
similarly, apart from the parameter $k + \delta$, with $\delta$ the degeneracy.

This paper is organized as follows. After an overview of related works, 
preliminary notions and results are given in Section~\ref{sec_preliminaries}.
Then, Section~\ref{sec_contractibility} proves our results regarding the contractibility
problem, and Section~\ref{sec_mcc} those on the maximum common contraction problem.

\medskip
\noindent
\textbf{Related works.}
A rich related line of work is the study of the \textsc{$H$-Contractibility}
problem on undirected, unlabeled graphs. It consists in deciding whether an
input graph $G$ can be transformed into a graph isomorphic to $H$ using only
edge contractions. In a seminal article~\cite{brouwer1987contractibility},
\textsc{$H$-contractibility} was proven NP-hard with $H=P_4$, the path on $4$
vertices. Follow-up works~\cite{levin2008computational,levin2008computationalb}
gave characterizations of the graphs $H$ such that \textsc{$H$-Contractibility}
is NP-hard. On the positive side, \textsc{$H$-Contractibility} was proven
polynomial-time solvable if $G$ is chordal~\cite{belmonte2012edge,belmonte2014detecting}, 
or if $H$ is planar (but
still fixed)~\cite{kaminski2010contractions}. A polynomial algorithm
for $H$ of bounded degree and $G$ of bounded treewidth was also given in~\cite{matouvsek1992complexity}, in addition to the proof that
deciding if a tree is a contraction of another is NP-hard (already mentioned above).
Variations of the problem, for instance deciding whether $k$ contractions
are enough to make a graph fall into a given \emph{class}, have also been studied,
including under the parameterized complexity point of view. These include the
contractibility into \emph{grid graphs}~\cite{saurabh2022parameterized},
\emph{bipartite graphs}~\cite{heggernes2013obtaining},
 \emph{paths/trees}~\cite{heggernes2014contracting,krithika2016lossy} or graphs of
\emph{bounded degree}~\cite{belmonte2014parameterized}.

As mentioned in the introduction, 
maximum common contractions were studied on \emph{phylogenetic
networks}~\cite{marchand2024finding}, which are directed acyclic graphs with a single root and labeled leaves.
In~\cite{marchand2024finding}, the NP-hardness of the problem is proven, and a polynomial-time tractable
sub-case is identified, but the parameterized complexity of the problem was not explored. 
Nonetheless, hardness proofs from~\cite{marchand2024finding} as well as~\cite{brouwer1987contractibility}
can be easily adapted to fully-labeled undirected graphs. It gives the following starting points: contractibility into a $P_4$ graph
is NP-hard on labeled graphs~\cite{brouwer1987contractibility},  
the maximum common contraction problem on labeled graphs is NP-hard if both
input graphs have constant degree (and thus also constant
degeneracy)~\cite[Theorem~8]{marchand2024finding},
and the labeled contractibility problem is NP-hard on graphs of bounded
degeneracy (also in \cite[Theorem~8]{marchand2024finding}, as although the
reduction
is to the maximum common contraction problem, it 
falls back to a contractibility instance of bounded degeneracy).

\section{Preliminary notions}
\label{sec_preliminaries}

For an integer $n$, we may use the notation $[n] = \{1,2,\ldots,n\}$.  We
denote the vertex set and edge set of a graph $G$ by $V(G)$ and $E(G)$,
respectively.  The subgraph of $G$ induced by a subset of vertices $X \subseteq
V(G)$ is denoted $G[X]$.  We write $G - X$ for the graph $G[V(G) \setminus X]$,
and if $X = \{x\}$ has a single element we may write $G - x$.  For a vertex
$u$, $N_G(u)$ is the set of neighbors of $u$ in $G$, and $N_G[u] = N_G(u) \cup
\{u\}$.  We say that two disjoint subsets $X, Y \subseteq V(G)$
are \emph{adjacent} if there exists an edge between an element of $X$ and an
element of $Y$.  The maximum degree of $G$ is denoted $\Delta(G)$.  The
\emph{degeneracy} of $G$ is $\delta(G)$, which is the smallest integer such
that every subgraph of $G$ has a vertex of degree at most $\delta(G)$.

Given two graphs $G$ and $H$, we write $G = H$ if and only if $V(G) = V(H)$ and
$E(G) = E(H)$.  This differs from the more common notion of
isomorphism: the vertex sets of the two graphs must consist of precisely the
same elements, and edges must connect the same pairs of vertices.  We can thus
view the vertices of $G$ and $H$ as uniquely labeled, and equality requires
that vertices with the same labels share the same edges in both graphs.
However, we prefer to avoid labeling functions, and instead compare
the vertices and edges directly. Note that sets of labels
may both intersect ($V(G)\cap V(H)\neq \emptyset$, in fact
no common contraction is possible otherwise) and contain elements not
present in the other graph ($V(G)\Delta V(H)\neq \emptyset$).

\noindent\textbf{Labeled contractions of graphs.} Given a graph $G$ and an edge
$uv$ of $G$, the \emph{labeled contraction} $(u,v)$ is an operation that
transforms $G$ as follows: (1) add an edge between $u$ and every vertex of
$N_G(v) \setminus N_G[u]$; (2) remove $v$ and all its incident edges.  The
graph obtained from $G$ after the labeled contraction $(u,v)$ is denoted
$G/(u,v)$.  Note that because the sets of vertices of our graphs matter,
contracting $(u, v)$ is different than contracting $(v, u)$, that is, $G/(u, v)
\neq G/(v, u)$ (as these two graphs remove a different vertex).  Do note that
$G/(u, v)$ and $G/(v, u)$ are isomorphic in the traditional sense. 

A \emph{labeled contraction sequence} $S$ on $G$ is a list of vertex
pairs $S = ((u_1, v_1), \ldots, (u_k, v_k))$ such that, for each $i \in \{0, 1,
\ldots, k - 1\}$, if $G_i$ is the graph obtained after the application of the
first $i$ labeled contractions (with $G_0 = G$), then $u_{i+1} v_{i+1}$ is an
edge of $G_i$ and $G_{i+1}$ is obtained by applying the labeled contraction
$(u_{i+1}, v_{i+1})$ on $G_i$.  We denote the resulting graph $G_k$ as $G/S$
(if $S$ cannot be applied on $G$, then $G/S$ is undefined).  The number of
pairs in $S$ is denoted $|S|$, which here is $k$.

We say that a graph $H$ is a \emph{labeled contraction of $G$} if there exists
a labeled contraction sequence $S$ such that $G/S = H$.  Again, we emphasize
that we require equality here.  If a graph $M$ is a labeled contraction of both
$G$ and $H$, it is called a \emph{common labeled contraction}.  A \emph{maximum
common labeled contraction} of $G$ and $H$ is a common labeled contraction with
a maximum number of vertices.  Equivalently, it is the result of applying a minimum number of labeled  contractions on $G$
and $H$. Formally,
we are interested in the following problem.  

\begin{mdframed}[nobreak=true]
\textsc{Maximum Common Labeled Contraction}\\
    \textbf{Input:} two graphs $G$ and $H$, integer $k$.\\
    \textbf{Question:} Are there labeled contraction sequences $S_1$ and $S_2$
    such that $G/S_1 = H/S_2$ and $|S_1|+|S_2|\leq k$ ?
\end{mdframed}
This can be viewed as the computation of a graph distance, since $|S_1| + |S_2|$ give a minimum
number of contractions and \emph{expansions} required to transform $G$ into
$H$, where expansions are the reverse of contractions. Note also that
the problem is equivalent to asking, given $G$ and $H$ whether
there is a common contraction $M$ such that $2|V(M)|\geq |V(G)|+|V(H)|-k$.  

An ``easier'' variant 
is when the common contraction must be the smaller input graph.

\begin{mdframed}[nobreak=true]
\textsc{Labeled Contractibility}\\
    \textbf{Input:} two graphs $G$ and $H$ with $V(H) \subseteq V(G)$.\\
    \textbf{Question:} is $H$ a labeled contraction of $G$?
\end{mdframed}

\begin{remark}
In the remainder, all contractions are labeled, so we may
simply write ``contraction'' instead of ``labeled contraction''.
\end{remark}

\maxcommonpb generalizes \contractibilitypb. Indeed,
$H$ is a contraction of $G$ if and only if $G$ and $H$ have a
common contraction of size at least $|V(H)|$, or using at most $|V(G)\setminus V(H)|$
contractions.
Therefore, our hardness results that apply to \contractibilitypb
transfer to \maxcommonpb.
Note that in either problem, we do not require the input graphs to be connected.  In \contractibilitypb, the connected components of $G$ can be matched uniquely to those of $H$ using common vertices. 
In \maxcommonpb, one could proceed as follows: for each connected component $C_1$ from $G$ and  $C_2$ from $H$, compute a maximum common contraction between $G[C_1]$ and $G[C_2]$, and assign a weight on  $\{C_1, C_2\}$ equal to the size of the common contraction ($-\infty$ if none exists).
Then, find a maximum-weight perfect matching in the resulting edge-weighted bipartite graph.  Hence, disconnected graphs only affect the complexity by a potential polynomial factor. 

\begin{figure}
    \centering
    \includegraphics[width=\textwidth]{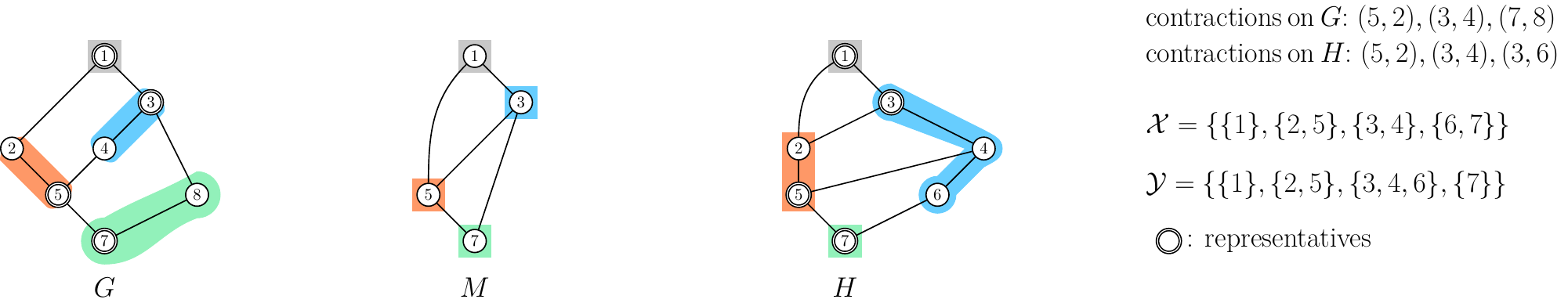}
    \caption{Example of two graphs $G,H$ and a common contraction $M$.
    On the graphs, the shaded areas represent the witness sets. The
    witness structures of respectively $G$ and $H$ into $M$ are called
    $\X$ and $\Y$, and are given on the right. Note that $V(G)\neq V(H)$,
    as allowed by our problem statement.}
    \label{fig_witness}
\end{figure}

\noindent\textbf{Witness structures.} When applying a contraction
sequence $G/S = H$, each vertex that gets removed from $G$ can be seen as
``subsumed'' by exactly one of the remaining vertices in $H$.  It is common to
replace the result of a contraction sequence with the vertex partition of
$V(G)$ that groups subsets of vertices that end up in the same vertex.  Such a partition \emph{witnesses} the existence of a contraction sequence
from $G$ to $H$, formalized as follows.

Let $G, H$ be two graphs with $V(H) \subseteq V(G)$.  Let $\W = \{W_1, \ldots,
W_{|V(H)|}\}$ be a partition of $V(G)$ with $|V(H)|$ non-empty sets.  For $u
\in V(G)$, we denote by $\W(u) \in \W$ the set of $\W$ that contains $u$.  We
say that $\W$ is a \emph{witness structure of $G$ into $H$} if all the
following conditions are satisfied (see Figure~\ref{fig_witness}):
\begin{itemize}
    \item 
    For every $W_i \in \W$, the induced subgraph $G[W_i]$ is connected.
    \item 
    Every $W_i \in \W$ contains exactly one vertex of $H$.  This vertex is called the \emph{representative} of $W_i$ (in $H$).
    \item 
    For every distinct $u, v \in V(H)$, $uv \in E(H)$ if and only if $\W(u)$ and $\W(v)$ are adjacent.
\end{itemize}
We note that there is a natural bijection between $\W$ and $V(H)$ formed by the
representatives.  The following equivalence is well-known in the case of
unlabeled graphs~\cite{brouwer1987contractibility,levin2008computational}, and is easily seen to hold in labeled graphs. An example of two graphs,
a common contraction, and the corresponding witness structures is given
on Figure~\ref{fig_witness}.

\begin{observation_rep}\label{obs:witness-structure}
    Let $G, H$ be graphs.  Then $H$ is a contraction of $G$ if and only if
    there exists a witness structure of $G$ into $H$.
\end{observation_rep}
\begin{proof}
The forward direction can be obtained by starting with $G$ and a partition of
$V(G)$ in which every vertex is by itself, then after each contraction $(u,
v)$, we update the set associated with $u$ by adding all the elements
associated with $v$.  When we are done, only $V(G)$ remains and the resulting
vertex-set association forms a witness structure.  For the reverse direction,
we note that the connected subgraphs $G[W_i]$ can all be contracted to a single
vertex independently, and under the conditions of witness structures, doing so
must result in $H$.
\end{proof}

Note that each contraction reduces the number of vertices by exactly one, and
so if $\W$ is a witness structure of $G$ into $H$, then the number of
contractions needed is $\sum_{W_i \in \W} (|W_i| - 1)$.  To finish, we prove
that contractions within a witness set can essentially be done in any order, as
long as the representative stays.  The idea is that if we apply any contraction
within a witness set first, we can update the affected witness set and get a
witness structure for the modified graph.

\begin{observation_rep}\label{obs:witness-order} Let $\W$ be a witness structure of
$G$ and $H$, and let $uv \in E(G)$ such that $u, v$ are in the same witness set
of $\W$.  If $v \notin V(H)$, then $H$ is a contraction of $G / (u, v)$.
\end{observation_rep}

\begin{proof} 
Denote $G' = G / (u, v)$.  It suffices to consider the partition
    $\W'$ of $V(G')$ obtained from $\W$ by just removing $v$ from $\W(u)$.  It
    is easy to see that $|\W'| = |V(H)|$, that $\W(u) \setminus \{v\}$ still
    contains the same representative of $V(H)$ since $v \notin V(H)$, and that
    $G'[\W(u)  \setminus \{v\}]$ is still connected.  Moreover, the edges with
    one endpoint in $\W(u)  \setminus \{v\}$ and the other outside are the same
    as $\W(u)$, and so $\W'$ is a witness structure of $G'$ into $H$.  By
    Observation~\ref{obs:witness-structure}, $H$ is a contraction of $G$.
\end{proof}

\section{Labeled Contractibility}
\label{sec_contractibility}

We begin by describing where the hardness barrier resides for the
contractibility problem.  We focus on difficult parameters first, then study
those that lead to FPT algorithms.

\subsection{W[1]-hardness in parameter $k$}

We first consider the parameter $k$, which is the number of contractions needed
to transform $G$ into $H$, or equivalently $k = |V(G)| - |V(H)|$.  We reduce
from the well-known \textsc{Multicolored Clique} problem, a W[1]-hard problem~\cite{fellows2009parameterized} which we recall. 

\begin{mdframed}
\textsc{Multicolored Clique}\\
    \textbf{Input:} a graph $G_C = (V, E)$ along with a partition $\{V_1,
    \ldots, V_k\}$ of $V$ into $k$ sets, with each $V_i$ called a color
    class.\\
    \textbf{Output:} does $G_C$ contains a \emph{multicolored clique}, i.e., a
    clique that contains exactly one vertex per color class?
\end{mdframed}

Let $G_C = (V, E)$ be an instance of \textsc{MultiColored Clique}, with $V =
V_1 \cup \ldots \cup V_k$ partitioned into $k$ color classes.  We assume
that for every distinct $i, j \in [k]$, each vertex $v \in V_i$ has at
least one non-neighbor in $V_j$.  This is without loss of generality, as we
can add an isolated vertex to each $V_i$ without changing the answer to the
instance.

Let us construct two graphs $G$ and $H$ from $G_C$.  We start with the simpler
$H$, which is obtained by copying $G_C$, and making each $V_i$ a
clique.  More specifically
\begin{align*}
V(H) = V(G_C) & & E(H) = E(G_C) \cup \bigcup_{i \in [k]} \{uv : u \in V_i, v \in V_i, u \neq v\}.
\end{align*}
To obtain $G$, start with a copy of $H$, then add $k$ new vertices $t_1,
\ldots, t_k$, which are not present in $H$.  Then, make $\{t_1, \ldots,
t_k\}$ a clique, and for each $i \in [k]$, add every possible edge between
$t_i$ and the vertices $V_i$ in $G$.  This concludes the construction of
$G$ and $H$.  

\begin{theorem_rep}
The \textsc{Labeled Contractibility} problem is W[1]-hard in parameter $k$,
which is the number of contractions needed to transform one input graph
into the other.    
\label{thm_contract_w1_k}
\end{theorem_rep}

\begin{proofsketch}
    Considering the above construction, if $G_C$ has a multicolored clique $u_1, \ldots, u_k$, then we can apply the set of contractions $(u_i, t_i)$, for $i \in [k]$.  This can only add edges between distinct $u_i, u_j$ vertices, but those edges were already present in $G$ (and in $H$) since the $u_i$'s form a clique.  In other words, this just gets rid of the $t_i$ vertices without adding any new extra edge, and thus the resulting graph is identical to $H$.

    Conversely, if $G$ can be contracted into $H$, then we can argue that each
    $t_i$ must be contracted with some $u_i \in V_i$ (contracting it with a
    $u_j \in V_j, j \neq i$ would add all edges between $u_j$ and $V_i$, which
    are not in $H$).  For $i\neq j$, if we contract
    $(u_i, t_i)$ then $(u_j, t_j)$, the edge $u_i u_j$ will be created if not
    already present.  This must be prevented to reach $H$, so $u_i u_j$ must
    already be in $G$, so the $u_i$'s contracted with the
    $t_i$'s must form a clique. 
\end{proofsketch}

\begin{proof}
Let $G_C$ be an instance of \textsc{Multicolored Clique} and let $G, H$ be constructed as described above.  We show that $G_C$ has a multicolored clique if and only if $H$ is a contraction of $G$.

Suppose that $G_C$ has a multicolored clique $u_1, \ldots, u_k$, where $u_i \in
V_i$ for each $i \in [k]$.  We present a witness structure $\W$ of $G$ into
$H$.  In $\W$, add the sets $\{u_i, t_i\}$ for each $i \in [k]$, and the
singleton sets $\{ \{v\} : v \in V(G_C) \setminus \{u_1, \ldots, u_k\} \}$.
This corresponds to the sequence of contractions $(u_1, t_1), \ldots, (u_k,
t_k)$.  It is clear that $|\W| = |V(H)|$, that each element of $\W$ has a
unique element of $V(H)$, and that each $G[W_i]$ is connected.  We must
argue that $vw \in E(H)$ if and only if there is some edge between $\W(v)$
and $\W(w)$ in $G$.

Let $v, w \in V(H)$ be two distinct vertices of $H$, which are also vertices of
$G$ and $G_C$.  If $vw \in E(H)$, then because $G$ started as a copy of
$H$, $vw \in E(G)$.  This edge still exists between $\W(v)$ and $\W(w)$ in
$G$, because $\W(v) \neq \W(w)$.  So instead assume that $vw \notin E(H)$.
Let $v \in V_i, w \in V_j$, and notice that $i \neq j$ since we made $V_i$
and $V_j$ cliques in $H$.  Suppose for contradiction that there is some
edge between $\W(u)$ and $\W(v)$ in $G$.  By construction, this edge cannot
be $vw$, and it cannot be $t_i w$ nor $t_j v$, which do not exist.  So
$\W(v)$ must contain $t_i$ and $\W(w)$ must contain $t_j$.  By the
construction of $\W$, this means that $v = u_i$ and $w = u_j$.  Since $u_i,
u_j$ are in a clique, $vw \in E(G)$ and $vw \in E(H)$, a contradiction.
Thus $\W(u)$ and $\W(v)$ are not adjacent in $G$, and it follows that $\W$
is a witness structure of $G$ into $H$.  By
Observation~\ref{obs:witness-structure}, $H$ is a contraction of $G$.

Conversely, suppose that one may apply at most $k$ contractions to $G$ to
obtain $H$.  Let $\W$ be a witness structure of $G$ into $H$.  For $i \in
[k]$, consider the vertex $t_i$ of $G$, and denote by $u$ the
representative of $\W(t_i)$, where $u \in V(H)$.  We claim that $u \in
V_i$, so assume instead that $u \in V_j$, with $i \neq j$.  Recall that
$t_i$ is adjacent to every vertex in $V_i$ in $G$, so if $u \in V_j$, then
$\W(u) = \W(t_i)$ is adjacent to every witness set that contains a vertex
of $V_i$.  But $u \in V_j$, a contradiction since we assume that $u$ has at
least one non-neighbor in $V_i$ in $G_C$, and thus in $H$ as well.  It
follows that the unique vertex $u$ of $G$ in the same witness set as $t_i$
belongs to $V_i$.  

Since this holds for every $i \in [k]$, this also implies that no two $t_i$
vertices belong to the same witness set.  We may then define the set of $k$
distinct vertices $u_1, \ldots, u_k$ where, for $i \in [k]$, $u_i$ is the
vertex of $V_i$ in the same witness set as $t_i$.  We argue that this set forms
a clique of $G_C$.  Let $i, j \in [k]$ with $i \neq j$.  Because of the edge
$t_i t_j$, $\W(t_i)$ and $\W(t_j)$ are adjacent in $G$.  Thus there is an edge
between $\W(u_i)$ and $\W(u_j)$, and by the conditions of witness sets the edge
$u_i u_j$ must exist in $H$.  Thus $u_1, \ldots, u_k$ forms a multicolored
clique of $G_C$.    
\end{proof}

We observe that this problem is easily seen to be in XP when parameterized by
$k$. Indeed,   
given graphs $G$ and $H$, we can just try every sequence of $k$
contractions in $G$ that suppress a vertex of $V(G) \setminus V(H)$.  There are $O(n^2)$ choices and we make at most $k$ contractions, resulting in complexity of the form $O(n^{2k})$. Note also that since our
reduction is from \textsc{Multicolored Clique} and it preserves the parameter $k$ exactly, it also implies~\cite[Corollary~14.23]{cygan2015parameterized} no $f(k)\cdot n^{o(k)}$ algorithm for \contractibilitypb under the Exponential-Time Hypothesis (ETH). Therefore,
the aforementioned XP algorithm is essentially optimal under ETH.

\subsection{para-NP-hardness for clique-width}

We recall the definition of clique-width.
Each vertex of a graph can be assigned a color, and the clique-width is the
minimum number of colors required to build a graph using a sequence of the
following operations:
\begin{itemize}
    \item Creation of a new vertex with color $i$.
    \item Disjoint union of two (vertex-colored) graphs.
    \item Adding all possible edges between vertices colored $i$ and
        vertices colored $j$,  $i\neq j$.
    \item Recoloring all vertices with color $i$ to color $j$, where $i\neq j$.
\end{itemize}
The clique-width of a graph $G$ is denoted $\cw(G)$.

We describe a reduction from {\sc Unary Bin Packing}, and more specifically the variant
in which we ask that every bin is filled exactly up to its capacity,
and the input integers are encoded in unary (an integer $a$ takes
$a$ bits in the input). We call it \sbinpacking.

\begin{mdframed}
\sbinpacking\\
    \textbf{Input:} Integers $a_0,\dots,a_{n-1}$ (item sizes) encoded in unary,
    bin capacity $C$, and number of bins $k$, such that $\sum_{x=0}^{n-1} a_x =
    Ck$.\\
    \textbf{Output:} Is there an assignment $\phi:[0,n-1]\rightarrow [0,k-1]$
    of each item to a bin such that $\forall i\in[0,k-1]$, $\sum_{j\text{ :
    }\phi(j)=i} a_j = C$ ?
\end{mdframed}

Note that item size  and bin subscripts are indexed at $0$ to simplify some
calculations later on.  We know from~\cite{jansen2013bin}
that {\sc Unary Bin Packing} is NP-hard and W[1]-hard in the number of bins $k$. It
turns out that \sbinpacking is also W[1]-hard in parameter $k$, and also
NP-hard~\cite{lafond2025cluster}. 
We use this problem to show that \textsc{Labeled Contractibility} is
NP-hard even when both input graphs have clique-width at most $4$.  

\begin{figure}
\centering
\includegraphics[width=\textwidth]{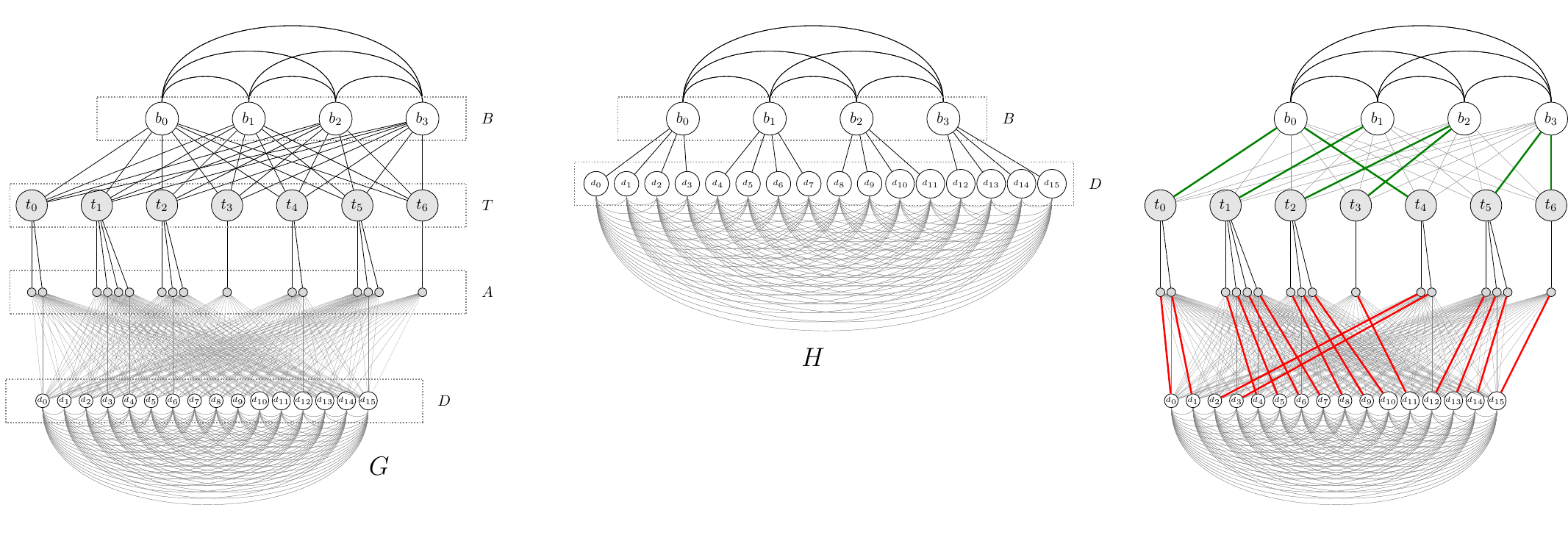}
    \caption{Illustration of the reduction from \sbinpacking to
    fully-labeled contractibility, on an instance. 
    This instance of \sbinpacking has $4$ bins (represented by the vertices
    labelled $b_0,\dots,b_3$), $7$ items (represented by the vertices $t_0,\dots
    t_6$), the items have size respectively $2,4,3,1,2,3$ and $1$ as represented
    by their number of neighbors (without counting $b_i$), and
    the size of each bin is $4$ (as represented by the number of neighbors each
    $b_i$ vertex has in $H$).
    On the right, the contraction of the highlighted edges yields $H$.
    As we discuss in the proof, the green edges then encode
    a solution to \sbinpacking
    }
    \label{clique_width_reduction}
\end{figure}

\noindent\textbf{Reduction description.} Consider an input to \sbinpacking,
with item sizes $a_0, \ldots, a_{n-1}$, bin capacity $C$, and number of bins
$k$.  An illustration of the reduction is given in
Figure~\ref{clique_width_reduction}.  From an instance $a_0,\dots,a_{n-1}, C,
k$ of \sbinpacking, we construct graphs $G,H$ as follows.
Both $G$ and $H$ contain a set of vertices $B = \{b_0, \ldots, b_{k-1}\}$, each
corresponding to a bin, connected as a clique in both
graphs.  Again in both $G$ and $H$, there is also a set of $C k$ vertices $D =
\{d_0, \ldots, d_{Ck - 1}\}$,  connected as a clique in both graphs. 
In $H$, each vertex $d_j \in D$ has exactly one neighbor in $B$, which is $b_i$
such that $\lfloor j/C\rfloor = i$ (thus each $b_i$ has exactly $C$
neighbors in $D$).
This finishes the construction of $H$.
In $G$, there also vertices $T = \{t_0, \ldots, t_{n-1}\}$ not present in $H$,
which represent the items.  The set $T$ forms an independent set.  The vertex
sets $T$ and $B$ form a bi-clique, so every possible edge between the two sets
is present.  For each $t_x \in T$, we also include in $G$ a set of vertices
$A_x = \{\alpha_x^0, \ldots, \alpha_x^{a_x - 1}\}$ containing $a_x$ elements,
each having $t_x$ as a neighbor.  We denote the set $A_0 \cup \ldots \cup
A_{n-1}$ as $A$ (see Figure~\ref{clique_width_reduction}), which forms an
independent set and is absent from $H$. The set $A$  forms a bi-clique with the
vertices $D$ in $G$.

For $H$ to be a contraction of $G$, there must be a way for each $b_i$
to acquire the neighbors specified by $H$. In particular,
each $b_i$ is connected in $H$ to exactly $C$ vertices of the set $D$ (recall
that $C$ is the bin capacity). Specifically, in $H$ there is an edge
$(b_i, d_j)$ for any $i\in [0,k-1]$ and $j\in [0,Ck-1]$ such that $\lfloor
j/C\rfloor=i$. 
To do this, each $b_i$ must be contracted with some $t_x$
vertices, making $b_i$ acquire a certain number of neighbors in $A$.  
Each such neighbor must then disappear, and to do that there must be 
a contraction with an element of $D$.  Since $|A| = |D|$, each element of $A$
is contracted with a single element of $D$, so $b_i$ must have acquired exactly
$C$ neighbors into $A$ to start with. We proceed to the proof of this idea.

\begin{theorem_rep}
The \textsc{Labeled Contractibility} problem is NP-hard, even if the input graphs have clique-width at most $4$.
\label{thm_cw}
\end{theorem_rep}

\begin{proofsketch}
    The graph $G$ actually has clique-width $3$. We can first create the $G[T
    \cup A]$ subgraph, a forest of star trees, using two colors.  We can create
    the $B$ clique and then connect it to all of $T$ using a third color,
    recoloring $B$ to the color of $T$, then adding $D$ and connecting it to
    $A$ reusing that third color.  As for $H$, we can incorporate each subgraph
    $H[\{b_i\} \cup (N_H(b_i) \cap D)]$ one at a time.  We assume we already
    have the $b_1, \ldots, b_{i-1}$ vertices using one color, and their
    neighbors in $D$ using another, and we add $b_i$ using a new color and its
    neighbors in $D$ using another new color, which allows connecting them to
    the previous $B$ and $D$ vertices (and then we recolor $b_i$ and its
    neighbors in $D$).

    To see that the generated instance is equivalent, suppose that $\phi$
    assigns items to bins perfectly.  If bin $i$ has items $a_{i_1}, \ldots,
    a_{i_q}$, in $G$ we contract $(b_i, t_{i_1}), \ldots, (b_i, t_{i_q})$.
    Since each $t_{i_j}$ has $a_{i_j}$ neighbors in $A$, this gives $b_i$
    exactly $C$ neighbors from $A$, say $v_0, \ldots, v_{C-1}$.  Then to ensure
    that $b_i$ has the correct neighbors from $D$, we contract $(v_0, d_{iC}),
    (v_1, d_{iC + 1}), \ldots, (v_{C-1}, d_{(i+1)C - 1})$.  This results in the
    graph $H$.

    Conversely, if we can contract $G$ into $H$, then we can argue in
    terms of witness sets: $b_i \in B$ cannot be in the same witness
    set as $v
    \in A$, since $b_i$ would become adjacent to all of $D$; a $t_i \in T$ cannot
    be with a $d_j \in D$, as $d_j$ would become adjacent to all of $B$.  This
    implies that each $t_i \in T$ is in the witness set of an element of $B$,
    and each $v \in A$ with an element of $D$.  Moreover, each $d_j \in D$
    \emph{must} have some $v \in A$ in its witness set, as otherwise $d_j$
    becomes impossible to connect with any $b_i \in B$.  Hence the witness sets
    containing vertices of $A$ and $D$ form a perfect matching.  This then
    makes it easy to show that if some $b_i$ has $t_{i_1}, \ldots, t_{i_q}$ in
    its witness set, then these members of $T$ must have had exactly
    $|N_H(b_i)| = C$ neighbors in $D$, establishing the correspondence with
    perfect bin assignments. 
\end{proofsketch}

\begin{proof} 
We first prove the correctness of the reduction,
    and then argue about the clique-width of the constructed graphs.

    Let $a_0,\dots,a_{n-1}, C, k$ be an instance of \sbinpacking, and $G,H$
    constructed as described above.  Suppose first that \sbinpacking is a
    yes-instance, i.e., that there exists an assignmnent $\phi$ of the items to
    the bins such that the sum of the sizes of the items in a bin is exactly
    $C$ for each of the $k$ bins. In $G$, we first contract each edge $(b_i,
    t_x)$ such that $\phi(x)=i$ (green edges on the example in
    Figure~\ref{clique_width_reduction}), keeping the vertex $b_i$.  Since each
    item is assigned to a bin, each $t_x$ is contracted into some vertex in
    $B$, therefore the set of vertices $T$ disappears. Since for each bin $i$,
    $\sum_{x\text{ : }\phi(x)=i} a_x = C$, and each $t_x$ has exactly $|A_x| =
    a_x$ neighbors in $A$, each $b_i$ now has exactly $C$ neighbors in $A$.
    Let $i$ be an integer between $0$ and $k-1$, and let us denote
    $v_i^{0},\dots,v_i^{C - 1}$ the $C$ neighbors of $b_i$ in $A$ in this
    partially contracted graph. We next contract each $v_i^y$ (for $0\leq y
    \leq C-1$) with the vertex  $d_{Ci+y}$ of $D$ (we can, as $A$ and $D$ form
    a bi-clique in $G$), keeping the $d_{Ci+y}$ vertices.  Doing this for all
    $i$, we delete the set $A$, which leaves only $B$ and $D$. As required, $B$
    $D$ each form a clique, and there is an edge $b_i d_j$ if and only if
    $\lfloor j/C\rfloor =i$.  We have therefore described a contraction
    sequence from $G$ to $H$.

    In the other direction, suppose that $H$ is a contraction of $G$.  Let $\W$
    be a witness structure of $G$ into $H$.  To ease notation a bit, we will
    denote by $W^b_i = \W(b_i)$ the witness set of each $b_i$ and by $W^d_j =
    \W(d_j)$ the witness set of each $d_j$. We argue that each element of $T$
    must be part of some $W^b_i$. Indeed, as $B$ and $D$ are the only vertex
    sets left in $H$, the sets $W^b_i$ and $W^d_j$ must form a partition of
    $V(G)$, and in particular of the elements of $A$ and $T$ are in one of
    those witness sets.  However, if a vertex $t_x$ from $T$ is in some
    $W^d_j$, then the after the contractions the vertex $d_j$ becomes connected
    to all of $B$, which is not the case in $H$.  Each vertex $t_x$ is
    therefore in one of the sets $W^b_i$. 
    
    We can then define $\phi$ as $\phi(x)=i$ if and only if $t_x\in W^b_i$
    (such an $i$ is unique as the $W^b_i$ do not intersect).
    
    Let us now argue that $\phi$ assigns items to bins perfectly.  Let $u$ be
    some vertex of $A$. If $u$ is in a witness set $W^b_i$, then after
    contracting the vertex $b_i$ becomes connected to all of $D$,
    which is not the case in $H$. Therefore, $u$ is in some
    witness set $W^d_j$. We also argue the opposite,
    that for any $d_j\in D$, there is some $u\in A$ such that 
    $u\in W^d_j$. Indeed, if some $W^d_j$ contains no element
    of $A$, then note that in $G$ the neighborhood of $d_j$ is composed
    of $D\setminus \{d_j\}\cup A$, and other elements of $D$ cannot be in $W^d_j$. 
    Therefore, we would have
    $W^d_j=\{d_j\}$.  In $G$, $\{d_j\}$ is only adjacent to witness sets that
    contain some vertex of $A$ or $D$.  As we just argued, each vertex of $A$
    is in the witness set of a $D$ vertex.  Thus the witness set $W^d_j$ 
    has no neighbor belonging to a $W^b_i$ set,
    which is a contradiction as $d_j$ shares an edge with some vertex of $B$ in
    $H$.
    
    Therefore, we have both that all vertices in $A$
    are in  some  $W^d_j$, and
    that each $W^d_j$ contains some vertex in $A$.
    Since $|A|=\sum_{x=0}^{n-1} a_x = Ck = |D|$,
    this implies that each $W^d_j$ is of the form $\{d_j,u\}$,
    for some $u\in A$ uniquely associated to $j$.
    In particular, this implies that
    each vertex $t_x$, which has $a_x$ neighbors in $A$,
    is therefore adjacent to exactly $a_x$ distinct
    witness sets $W^d_j$. We write $S_x=
    \{d_j \in D:  N_G(t_x)\cap W^d_j\neq \emptyset\}$, i.e., the set of vertices
    of $D$ whose witness set is adjacent to that of $t_x$. 
    We have $|S_x|=a_x$, and for $x\neq y$, we also
    have $S_x\cap S_y=\emptyset$. Now, $\forall i\in[0,k-1]$, 
    $\cup_{x\text{:}\phi(x)=i} S_x$ denotes the representatives 
    of the witness sets of $D$ neighbor to $W^b_i$, and therefore
    the neighborhood of $b_i$ outside of $B$ in $H$.
    As each $b_i$ has exactly $C$ neighbors outside of $B$ in $H$,
    we must have $\sum_{x\text{ : }\phi(x)=i} |S_x| = \sum_{x\text{ :
    }\phi(x)=i} a_x=C$. Therefore, $a_0,\dots,a_{n-1},C,k$ is a yes-instance.

    \noindent\textbf{Clique-width analysis.} We finish by arguing that both $G$
    and $H$ have clique-width at most $4$. Let us start with $H$, which we
    construct inductively on $k = 0, 1, \ldots, k - 1$.  For $k=0$, we add
    vertices $d_0,\dots,d_{C-1}$ with color $1$, $b_0$ with color $2$, and add
    all edges between colors $1$ and $2$. This finishes the construction for
    $k=0$, with vertices of $D$ added so far having color $1$ and those of $B$
    having color $2$.  Then, given the graph $H$ constructed with $k-1$
    vertices in $B$, with all vertices of $D$ having color $1$ and those of $B$
    having color $2$, we introduce the vertices $d_{(k-1)C},\dots,d_{kC-1}$
    under color $3$, vertex $b_{k-1}$ under color $4$.  We add all edges
    between color $3$ and $4$ (which creates exactly the required edges between
    $b_{k-1}$ and $d_{(k-1)C},\dots,d_{kC-1}$), then connect color $1$ and $3$
    as a bi-clique (making $D$ a clique) and likewise for colors $2$ and $4$
    (making $B$ a clique). We finish this step by recoloring $3$ with $1$ and
    $4$ with $2$, in case we may pursue the construction.

    As for $G$, it actually has clique-width $3$.   We can first construct the
    subgraph induced by $A$ and $T$ using two colors.  Indeed, $G[A \cup T]$ is
    a forest of star trees, so we can just create each star tree independently
    with two colors, and then take the disjoint union of all those star trees.
    This can be done in a way that $G[A \cup T]$ is built with vertices of $A$
    having color $1$ and those of $T$ having color $2$.  We then construct the
    $B$ clique independently using two colors and recolor every vertex of $B$
    to $3$. We then add $B$ to the construction and add all edges between $B$
    and $T$ (colors $1$ and $3$).  Then recolor $3$ to $1$, so that $B$ and $T$
    have the same colors.  In the same manner, we construct the $D$ clique and
    give it color $3$, and add edges between colors $2$ and $3$ ($A$ and $D$).
    This results in $G$.
\end{proof}

\subsection{An FPT algorithm in $k$ and degeneracy}

Let us start with a definition of the degeneracy. Given a total order $\sigma$
of the vertices of a graph $G$, we denote by $N_{\geq,\sigma}^G(u)$ the neighbors
of $u$ that are after it according to $\sigma$.  We define the degeneracy $\delta(G)$
of a graph as:
$\delta(G) = \min_{\sigma}\max_{u\in V(G)} |N_{\geq,\sigma}^G(u)|$.

We call $N_{\geq,\sigma}^G(u)$ the \emph{remaining neighborhood} of $u$ according to $\sigma$.
This denomination is to be understood in an interpretation of degeneracy as an
elimination process. 
\begin{lemma_rep}
    If $H$ is obtained from $G$ by $\leq k$ contractions, then $\delta(H)\leq \delta(G)+k$
    \label{lemma_degen_contraction}
\end{lemma_rep}
\begin{proofsketch}
    We prove that a single contraction of an edge $u,v$ in a graph only
    increases the degeneracy by at most one. Given an elimination ordering
    $\sigma$ optimal for $\delta(G)$, we modify it to place the merged vertex
    (whose degree might have doubled compared to $u,v$) as late into $\sigma$
    as possible. By placing it after all of its neighbors, we make
    its remaining neighborhood empty, while augmenting it by
    at most $1$ vertex for other vertices.
\end{proofsketch}
\begin{proof}
We prove the result by induction. The base case $k=0$ is trivial.
    Let us therefore consider now $G'$
    obtained from $G$ by a single contraction, of $u$ into $v$. Let also
    $\sigma$ be an elimination order such that $\max_{x\in V(G)}|N_{\geq,\sigma}^G(x)|=\delta(G)$. We distinguish below two cases. In both, we find an order
    $\sigma'$ of the vertices of $G'$ such that 
    $\max_{x\in V(G')}|N_{\geq,\sigma'}^{G'}(x)|\leq \delta(G)+1$, which allows
    to conclude.
    \begin{itemize}
        \item If $N(u)\cup N(v)$ (neighborhood of $v$ in $G'$) contains no element
        ranked higher than $v$ in $\sigma$, we simply get $\sigma'$ by removing $u$
        from $\sigma$. We have $N_{\sigma,\geq}^G(v)=N_{\geq,\sigma'}^{G'}(v)=\emptyset$.
        As for $x\in V(G')$ and $x\neq v$, $|N_{\geq,\sigma'}^{G'}(x)| \leq |N_{\geq,\sigma}^G(x)|$ if $x$ is before $u$ in $\sigma$ (the size decreases by $1$
        if $x$ is a neighbor of both $u$ and $v$ and stays the same otherwise), 
        $|N_{\geq,\sigma'}^{G'}(x)| = |N_{\geq,\sigma}^G(x)|$ if $x$ is after $v$
        in $\sigma$ and $|N_{\geq,\sigma'}^{G'}(x)| \leq |N_{\geq,\sigma}^G(x)|+1$
        if $x$ is between $u$ and $v$ in $\sigma$ (with an increase in the case where
        $x$ is a neighbor of $u$ but not $v$). Overall, we do have 
        $\max_{x\in V(G')}|N_{\geq,\sigma'}^{G'}(x)|\leq \delta(G)+1$.
        \item Otherwise, let $t$ be the highest-ranked element of $N(u)\cup N(v)$
        by $\sigma$ (it is after $v$, as otherwise we would be in the previous case).
        In this case, we get $\sigma'$ by removing $u$ an moving $v$ right after
        $t$ in $\sigma$. By doing so, we have $|N_{\geq,\sigma'}(v)|=0$, while
        for $x\in V(G')$ such that $x\neq v$,  
        $|N_{\geq,\sigma'}^{G'}(x)| \leq |N_{\geq,\sigma}^G(x)|+1$ if $x$ is a neighbor
        of $u$ after $u$ in $\sigma$, or if $x$ is a neighbor of $v$ after $v$ in $\sigma$.
        If $x$ is not a neighbor of $u$ and $v$, or if $x$ is before $u$ in $\sigma$,
        $|N_{\geq,\sigma'}^{G'}(x)| \leq |N_{\geq,\sigma}^G(x)|$. Again, 
        we obtain 
        $\max_{x\in V(G')}|N_{\geq,\sigma'}^{G'}(x)|\leq \delta(G)+1$.
    \end{itemize} 
    We have therefore found an ordering $\sigma'$ of $G'$ such that
        $\max_{x\in V(G')}|N_{\geq,\sigma'}^{G'}(x)|\leq \delta(G)+1$.
        By definition of the degeneracy, $\delta(G')\leq \delta(G)+1$.
        Therefore, by induction over the number $k$ of contractions, $\delta(H)\leq \delta(G)+k$
\end{proof}

We describe a branching algorithm for \contractibilitypb 
that tries a bounded number of contractions at
each recursion.  On a recursive call that receives graphs $G, H$ and integer
$k$, the enumeration proceeds as described below.
This algorithm is also given in pseudo-code in the appendix. 
\begin{itemize}
    \item If $G=H$ we return true. If $k<0$, or  $|V(G) \setminus V(H)| > k$, we return false.

    \item Otherwise, we take a vertex $u$ of $G$ of minimum degree.  Assuming
    that $G$ is the result of at most $k$ contractions, by
Lemma~\ref{lemma_degen_contraction}, the degree of $u$ is at most
    $\delta(G)+k$.

    \item 
    If $u$ is in $G$ but not in $H$, we know that $u$ must be removed by a contraction at some
    point.  So we branch into at most $\delta(G)+k$ ways of contracting $u$ with
    one of its neighbors (which is safe by Observation~\ref{obs:witness-order}
    on the order of contractions),
    and decrease $k$ by $1$.  

    \item 
    Otherwise, $u$ is in both $G$ and $H$ and must be kept.  If $N_G(u) = N_H(u)$, then we may simply ignore $u$ from now on, as further contractions on $G$ that remove vertices outside of $H$ will not change the neighborhood of $u$.  We thus remove $u$ from both graphs (note that this step may fail on the maximum common contraction problem).  
    \item If $N_G(u) \subsetneq N_H(u)$, i.e.\ 
    $u$ has a neighbor $v$ in $H$, but $v \notin N_G(u)$, then we know that
    some contraction must affect $u$ or $v$, as otherwise the edge $uv$ will
    never be created.  We branch into all ways of contracting $u$ with one of
    its neighbors not in $H$, or $v$ with one of its neighbors not in $H$.  
    (if there is no such neighbor for either of them, we return false).
    This
    branches into at most $2k$ cases,
    because $|V(G)\setminus V(H)|\leq k$.

    \item 
    If $u$ has a neighbor $v\in V(H)$ in $G$, but $v \notin N_H(u)$ 
    (i.e.\ both $u,v$ must be kept, but they share an unwanted edge) then we
    return false.

    \item 
    Finally, if none of the above holds, then $N_H(u) \subseteq N_G(u)$, and $u$ has a neighbor $v$ in $G$ with $v \notin V(H)$.  
    This vertex $v$ must be contracted into one of its neighbors.    
    It cannot be contracted into a vertex of $V(H)\setminus N_H(u)$ 
    (as $N_H(u)$ is
    already complete), it may therefore only be contracted into a vertex in
    $V(G)\setminus V(H)$(which contains $\leq k-1$ elements when not counting $v$)
        or $N_G(u)\cup\{u\}$ (which contains $\leq \delta(G)+k+1$ elements) 
    We branch
    over these $\delta(G)+2k$
    possibilities.

    
\end{itemize}


\begin{toappendix}
    
\begin{algorithm}
    \SetKwProg{Fn}{Function}{}{end}
    \SetKwData{return}{return}
    \SetKwFunction{contractibility}{\sf contractibility}

    \Fn{\contractibility{$G,H,k$}}{
    \If{$G=H$}{\return\textsf{true}\;}
    \If{$k<0$}{\return \textsf{false}\;}
    \If{$V(H)\not\subset V(G)$ or $|V(G)\setminus V(H)|> k$}{\return\textsf{false}\;}
    Let $u$ be a vertex of $G$ of minimum degree\;
    \If{$u\in V(G)\setminus V(H)$}{
        \For{$v\in N_G(u)$}{
            \If{\contractibility{$G/(u,v),H,k$}}{
                \return \textsf{true}\;
                }
            }
        \return \textsf{false}\;
        }
    \If{$u\in V(H)$}{
        \If{$N_G(u)=N_H(u)$}{\return \contractibility{$G\setminus\{u\},H\setminus\{u\},k$}}
        \If{$N_H(u)\not\subset N_G(u)$}
        {
            \For{$v\in V(H)$ s.t.\,$(u,v)\in E(H)\setminus E(G)$}{
                \For{$w\in N_G(v)\setminus V(H)$}{
                    \If{\contractibility{$G/(v,w),H,k-1$}}{
                        \return \textsf{true}\;
                    }
                }    
                \For{$w\in N_G(u)\setminus V(H)$}{
                    \If{\contractibility{$G/(u,w),H,k-1$}}{
                        \return \textsf{true}\;
                    }
                }
            }
            \return \textsf{false}\;
        }
        \If{$N_H(u)\subset N_G(u)$ and $\exists v\in N_G(u)\setminus V(H)$}{
            Let $v\in N_G(u)\setminus V(H)$\;
            \If{$N_G(v)\cap (V(H)\setminus N_H(u))=\emptyset$}
            {
                \If{\contractibility{$G/(u,v),H,k-1$}}{
                \return \textsf{true}\;
                }
            }
            \For{$w\in N_G(v)\setminus (V(H)\setminus N_H(u))$}{
                \If{\contractibility{$G/(v,w),H,k-1$}}{
                \return \textsf{true}\;
                }
            }
            \return \textsf{false}\;
        }
    }
    }
    \caption{FPT algorithm for $\delta+k$.}
    \label{algo:degen}
\end{algorithm}

\end{toappendix}

\begin{theorem_rep}
    \contractibilitypb can be solved in time $O\left((\delta(G)+2k)^k\cdot
    (n+m)\right)$, where $k$ is the number of contractions, $n=|V(G)|$, and
    $m=|E(G)|$.
    \label{thm:degen_k}
\end{theorem_rep}
\begin{proofsketch}
    The algorithm is a bounded search tree with two kinds of recursive calls:
    non-branching calls (when $N_H(u)=N_G(u)$) and branching calls
    involving a contraction.  In the latter, we make at most $\delta(G) + 2k$ recursive calls that each decrease
    $k$ by $1$.
\end{proofsketch}
\begin{proof}
    \textbf{Correctness:} We prove the correctness of the bounded search tree
    approach informally described above, and whose pseudocode is available
    in Appendix, by induction over the number of contractions. If $k=0$,
    an equality check is enough, and is indeed carried out.
    Assuming that search is correct up to $k-1$ contractions,
    consider $G,H,k$ an input to the recursion, and $u$ of minimum
    degree in $G$. 
    We will consider also $\W$, a witness structure of $G$ into $H$
    in the case where $G$ is contractible to $H$.
    To start with, either $u\in V(H)$ or not. In the former
    case, one cannot obtain $H$ if none of the edge adjacent to $u$ is contracted.
    In the latter ($u\in V(H)$), either $u$ already has exactly the correct
    set of neighbors or not. It it has the correct set of neighbors,
    we can simply remove it. Indeed, we have then that $G$ is contractible
    into $H$ if and only if $G\setminus\{u\}$ is contractible into $H\setminus\{u\}$ (which can easily be seen with a witness structure).
    If it does not have the correct set of neighbors, $u$ either (1)
    misses an edge with $v\in V(H)$, (2) has an extra neighbor $w\in V(H)$ not
    in $N_H(u)$ or (3) has the right set of neighbors from $V(H)$ but is
    also neighbor to vertices in $V(G)\setminus V(H)$. In case (1),
    if $H$ is a contraction of $G$, then the witness sets $\W(u)$
    and $\W(v)$, which are connected in $G$, must be adjacent. This
    implies at least one contraction incident to $u$ or $v$. In case (2),
    no contraction can both keep $u$ and $w$ while deleting their edge 
    ($\W(u)$ and $\W(w)$ are forced to be adjacent), so we correctly return
    false. As for (3), we note that a neighbor $v$ of $u$ in this situation
    must be part of some witness set $\W(y)$. If $y$ is not a neighbor
    of $u$, then $v\in \W(y)$ would introduce an unwanted edge. 
    Therefore, either $y=u$ or $y\in N_H(u)$, and $v$ may be connected
    to $y$ in $\W(y)$ either directly by an edge, or by a path of vertices
    in $V(G)\setminus V(H)$. Both of these cases are enumerated by the algorithm.
    In all cases leading to a contraction, Observation~\ref{obs:witness-order}
    allows to apply it first and recurse.

    \textbf{Complexity} A graph given to a recursive call
    is the result of a series of vertex deletions (of a vertex
    of minimum degree) and edge contractions
    compared with the original graph $G$. By definition, removing
    a vertex of minimum degree does not increase the degeneracy of a graph.
    As for the contraction, each one increases the degeneracy by at most
    $1$ (Lemma~\ref{lemma_degen_contraction}). The number of recursive
    calls involving a contraction is at most $k$, hence
    a degeneracy of the input graph bounded by $\delta(G)+k$. These
    recursive calls are the only ones involving \emph{branching} over
    several possibilities, at most $\delta(G)+2k$ of them, hence
    a recursive tree of size upper-bounded by $(\delta(G)+2k)^k$.
    To finish, equality over fully-labeled graphs may be tested in $|V(G)|+
    |E(G)|$.
    \textbf{Complexity analysis} Each branching recursive call (a branching call
    is one that gives rise to more than one call) reduces $k$ by $1$. 
    In the case where the vertex of minimum degree $u$ is in $V(G)\setminus V(H)$,
    the number of recursive calls is $\leq \delta(G)+k$ (the degree of $u$). In
    the case where $u\in V(H)$, and $N_H(u)\not\subset N_G(u)$, there
    are $2k$ recursive calls. In the other cases, there at most $k$ recursive calls.
    This gives an upper bound of $\max(2k,\delta+k)^k$ recursive calls.
    In each of these, the equality test takes $O(n^2)$.
\end{proof}

\section{Maximum Common Labeled Contractions}
\label{sec_mcc}

We now turn to maximum common contractions.  We start with our hardness result
on graphs of small degeneracy, contrasting with the positive result for the
contractibility problem.

\subsection{W[1]-hardness in 4-degenerate graphs}

We show that \textsc{Maximum Common Labeled Contraction} is W[1]-hard in
parameter $k$ even on graphs of degeneracy $4$.  Therefore, there is little hope for an FPT algorithm in parameter $k + \delta(G)$.  To give credit where is due, the reduction was initially inspired by~\cite{meybodi2020parameterized} for the hardness of domination problems on graphs of bounded degeneracy, although the  adaptations required for our proof makes it largely different. 
We will need the following lemma, which is a generalization of Observation~\ref{obs:witness-order}.

\begin{lemma_rep}\label{lem:matching}
    Let $H$ be a contraction of a graph $G$, and let $\W$ be the witness
    structure of $G$ into $H$.  Let $R \subseteq E(G)$ be a matching of $G$
    such that, for every edge $xy \in R$, the vertices $x$ and $y$ are in the
    same witness set of $\W$.

    Let $G'$ be the graph obtained after contracting every edge of $R$, making
    sure not to remove a representative vertex of $\W$.  Then $H$ is a
    contraction of $G'$.
\end{lemma_rep}

\begin{proof}
    As in the proof of Observation~\ref{obs:witness-order}, it suffices to
    observe that if we start from $\W$ and contract $(x, y)$, where $xy \in R$,
    then after removing $y$ from $\W(y)$, the resulting collection $\W'$ is a
    witness structure of $G / (x, y)$ into $H$ (this of course requires $y$ to
    not be a representative).  Moreover, if $R \setminus \{xy\}$ is non-empty,
    then any edge that remains still has its ends in one witness set, because
    $R$ is a matching.  Thus we may continue applying such contractions
    inductively.  
\end{proof}

\begin{figure}
    \centering
    \includegraphics[width=1\linewidth]
    {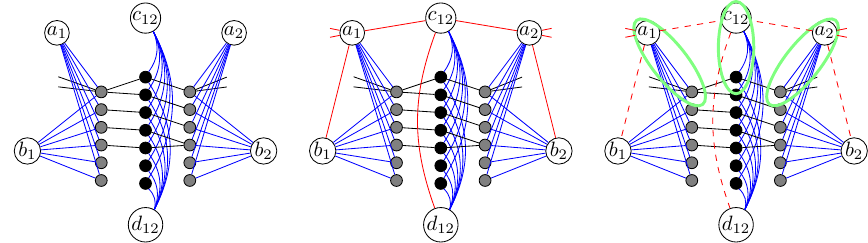}
    \caption{Left: an illustration of the vertices of and edges of $G$ relating
    $V_1$ and $V_2$.  Gray vertices represent $V_1'$ and $V_2'$, and black
    vertices represent $Z'_{12}$ (the bottom two vertices in each group
    represent extra vertices). Some edges are colored blue only for clarity.
    Middle:  the same subgraph but in $H$.  The red edges are those that are
    unique to $H$.  Note that $a_1$ has other neighbors $c_{13}, \ldots,
    c_{1k}$, similar with $a_2$, but all the neighbors of $b_1, b_2, c_{12},
    d_{12}$ are shown.  Right: witness sets corresponding to contractions in a
    solution for the forward direction (assume witness sets $\{a_1, u_1\},
    \{a_2, u_2\}, \{c_{12}, z_e\}$ such that $e = u_i u_j$).  Whether the red
    edges are present or not, this will result in the same graph}
    \label{fig:contrdegen}
\end{figure}

We can now describe our reduction.  Let $G_C = (V, E)$ be an instance of
\textsc{Multicolored Clique} with $V = V_1 \cup V_2 \cup \ldots \cup V_k$
partitioned into $k$ color classes.  Construct an instance $G, H$ of
\textsc{Maximum Common Labeled Contraction}, starting with the description of
$G$.  The reader may refer to Figure~\ref{fig:contrdegen}.  First, for each $i
\in [k]$, define $V_i'$ as a set of vertices obtained by starting with $V_i$,
then adding $4k^3$ new vertices.  We refer to elements of $V_i' \setminus V_i$
as \emph{extra vertices}.  Add to $G$ the set of vertices $V_1' \cup \ldots
\cup V_k'$.  Then for each distinct $i, j \in [k]$, and for each edge edge $e =
uv$ of $G_C$ with $u \in V_i$ and $v \in V_j$, create a new vertex $z_e$ and
add the path $u - z_e - v$ (or said differently, add the edge $uv$ to $G$ and
subdivide it, which creates $z_e$).  We may write $z_e = z_{uv} = z_{vu}$.
Denote by $Z_{ij}$ the set of $z_e$ vertices  created at this step (we consider
that $Z_{ij} = Z_{ji}$).  Next, obtain the set of vertices $Z'_{ij} = Z'_{ji}$
by adding $4k^3$ new vertices to $Z_{ij}$ (these new vertices are also called
extra vertices).

Then for each $i \in [k]$, add two new vertices $a_i$ and $b_i$, and make $a_i$
and $b_i$ adjacent to every vertex of $V'_i$ in $G$.  Note that $a_i$ and $b_i$
are not adjacent, and that the extra vertices of $V'_i$ are only adjacent to
$a_i$ and $b_i$.  Finally, for each distinct $i, j \in [k]$, add two vertices
$c_{ij}$ and $d_{ij}$, non-adjacent.  We define $c_{ji} = c_{ij}$ and $d_{ji} =
d_{ij}$.  Make $c_{ij}$ and $d_{ij}$ adjacent to every vertex in $Z'_{ij}$.  In
particular, extra vertices of $Z'_{ij}$ are only adjacent to $c_{ij}, d_{ij}$.
This completes the construction of $G$.
To construct $H$, start with a copy of $G$, then add the following edges:
$a_ib_i$ for each $i \in [k]$; $a_i c_{ij}$, $a_j c_{ij}$ and $c_{ij} d_{ij}$
for each distinct
$i, j \in [k]$;
Finally, define $K := 2(k + {k \choose 2})$.
This concludes the construction.

\begin{toappendix}
We now present the full proof of hardness of the maximum common contraction problem on parameter $k$, on graphs of bounded degeneracy.  
For clarity, for the construction required for Theorem~\ref{thm:4-degen} from an instance $G_C$ of the multicolored clique problem, we include the full set of vertices and edges here:
\begin{align*}
    V(G) &= V(H) = \bigcup_{i \in [k]} (V'_i \cup \{a_i, b_i\}) \cup \bigcup_{1
    \leq i < j \leq k} (Z'_{ij} \cup \{c_{ij}, d_{ij}\}) \\
    E(G) &= E_1 \cup E_2 \cup E_3, \mbox{where} \\
    E_1 &= \bigcup_{i \in [k]} \{a_i u : u \in V'_i\}  \cup \{b_i u : u \in
    V'_i\}\\
    E_2 &= \bigcup_{1 \leq i < j \leq k} \{c_{ij} z : z \in Z'_{ij} \} \cup
    \{d_{ij} z : z \in Z'_{ij} \} \\
    E_3 &= \bigcup_{e = uv \in E(G_C)} \{u z_{uv}, v z_{uv}\} \\
    E(H) &= E(G) \cup \bigcup_{i \in [k]} \{a_i b_i\} \cup \bigcup_{1 \leq i < j \leq k} \{c_{ij} d_{ij}, c_{ij} a_i, c_{ij} a_j\}.
\end{align*}

The idea is that $G$ must ``acquire'' the edges of $E(H) \setminus E(G)$ if it
wants to reach a common contraction (red in the figure).  One way to do so is
that each $a_i$ chooses a $u_i \in V_i$ to contract with, and each $c_{ij}$
chooses a $z_e \in Z_{ij}$.  The edge between $a_i$ and $c_{ij}$, and between
$a_j$ and $c_{ij}$, will be created in $G$ only if the chosen $u_i$ and $u_j$
form an edge $e$, and $c_{ij}$ was contracted with $z_{u_iu_j}$, as in the
figure.  We want this behavior to happen for every $i, j$ --- the $b_i$ and
$d_{ij}$ vertices, the extra vertices, and $K$ are setup to so that this is
feasible if and only if $G_C$ has a multicolored clique.  
\end{toappendix}

\begin{theorem_rep} \label{thm:4-degen}
    The \textsc{Maximum Common Labeled Contraction} is W[1]-hard in parameter
    $k$, the total number of contractions needed to achieve a common
    contraction, even if both input graphs have degeneracy $4$.  
\end{theorem_rep}

\begin{proofsketch}
    The graphs $G, H$ have degeneracy at most $4$ since we can have an elimination order that removes every $Z'_{ij}$ first (vertices of degree $4$), then the $V'_i$ vertices (remaining degree $2$), followed by the $c_{ij}, d_{ij}$ vertices and then the $a_i$ and $b_i$ vertices.

    Next, if $G_C$ has a multicolored clique $u_1, \ldots, u_k$, in either graph $G$ or $H$, we contract each $(a_i, u_i)$ pair, and each $(c_{ij}, z_{u_iu_j})$ pair, as in Figure~\ref{fig:contrdegen}.  Since, in $G$ and $H$, the edges $u_i z_{u_i u_j}$ and $u_j z_{u_i u_j}$ exist, the effect is 
    ``collapsing'' these red edges in $H$.
 Since the rest of the graphs are the same, this results in identical graphs.  The other direction assumes that $K$ contraction suffice on $G$ and $H$ to reach a common contraction, and produces a multicolored clique of $G_C$. 
 We can argue that in $G$, one of $a_i$ or $b_i$ is incident to a contraction
 because of the red edge $a_i b_i$ in $H$, and one of $c_{ij}$ or $d_{ij}$ is
 incident to a contraction because of $c_{ij} d_{ij} \in E(H)$.  These require
 $K/2$ contractions in $G$ and there must be $K/2$ contractions in $H$.
 These are 
 all of the contractions, which are therefore a matching.
 Lemma~\ref{lem:matching} is used in a more involved argument  to show that it
 is actually the $a_i$ and $c_{ij}$ vertices that are incident to a contraction
 (using $b_i$ or $d_{ij}$ vertices leads us to add too many new neighbors to
 extra vertices).  Once this is established, we can argue that the contraction
 partners $u_i$ of $a_i$ and $u_j$ of $a_j$ must form an edge $u_i u_j$, because
 the situation from Figure~\ref{fig:contrdegen} on the right is enforced. 
\end{proofsketch}

\begin{toappendix}

\begin{proof}[Proof of Theorem~\ref{thm:4-degen}]
Consider an instance $G_C = (V, E)$ of \textsc{Multicolored Clique} and graphs
    $G, H$ constructed as above, along with parameter $K = 2(k + {k \choose
    2})$, the total number of contractions.  

We first argue that the degeneracy of $G$ and $H$ is at most 4.  Since $G$ is a
    subgraph of $H$, it suffices to consider $H$ only.  We provide an
    elimination sequence in which every vertex has at most four neighbors
    before being deleted.  First consider any extra vertex in a $V'_i$ or
    $Z'_{ij}$ set.  These have degree two, so we may delete them first.  Next
    consider any $z_e$ vertex of $H$ from some $Z_{ij}$ set, where $e = uv$.
    It has four neighbors $\{u, v, c_{ij}, d_{ij}\}$.  Delete all the $Z_{ij}$
    vertices, for every $i, j$.  After that, consider any $u$ vertex in some
    $V_i$ set.  It has two remaining neighbors $a_i$ and $b_i$.  Delete all of
    those next.  Then delete the $d_{ij}$ vertices (one remaining neighbor
    $c_{ij}$) and the $c_{ij}$ vertices (two remaining neighbors $a_i, a_j$).
    The resulting graph has maximum degree $1$ as the $a_i b_i$'s form a
    matching, and it follows that $H$ and $G$ have degeneracy $4$.

We next show that $G_C$ admits a multicolored clique if and only if $G, H$ have
    a common contraction that can be achieved with a total of at most $K$
    contractions.

Suppose that $G_C$ admits a multicolored clique $u_1, u_2, \ldots, u_k$, where
    each $u_i$ is in $V_i$.  We perform the same contractions in $G$ and $H$.
    In either graph:\\ - for each $i \in [k]$, contract $(a_i, u_i)$;\\ - for
    each distinct $i, j \in [k]$, contract $(c_{ij}, z_{u_i u_j})$.

Note that the kept vertex is $a_i$ and $c_{ij}$, respectively.
This clearly requires $K = 2(k + {k \choose 2})$ contractions in total.  We must argue that applying them in $G$ or $H$ results in the same graph.
Let $M_G$ and $M_H$ be the graphs obtained by applying the above contractions on $G$ and $H$, respectively.
We use the witnesses for our arguments.  

Let $\W$ be the witness structure of $G$ into $M_G$.  
Note that $\W$ contains $\{a_i, u_i\}$ for $i \in [k]$, $\{c_{ij}, z_{u_i u_j}\}$ for distinct $i, j \in [k]$, and every other vertex is in a witness set of size $1$.  
Also note that $\W$ is also a witness structure of $H$ into $M_H$, since we apply the same contractions on both graphs.  
We argue that two witness sets of $\W$ are adjacent in $G$ if and only if the same two witness sets are adjacent in $H$, from which we can deduce that $M_G = M_H$.  

\begin{itemize}
    \item First consider $i \in [k]$ and some vertex $v \in V'_i \setminus
        \{u_i\}$.  Then $\{v\}$ is an element of $\W$.  From the construction,
        we see that $H$ adds no neighbor to $v$.  Thus $N_G(v) = N_H(v)$, and
        it follows that $\{v\}$ is adjacent to some $W \in \W$ in $G$ if and
        only if $\{v\}$ is adjacent to $W$ in $H$.

    \item Consider distinct $i, j \in [k]$ and a $z \in Z'_{ij} \setminus
        \{z_{u_i u_j}\}$ vertex.  Then $\{z\} \in \W$, and again from $N_G(z) =
        N_H(z)$ we make the same conclusion as in the previous case.

    \item From now on we do not need to consider the adjacencies of $\{v\}$ and
        $\{z\}$ witness sets.  Only witness sets of the form $\{b_i\},
        \{d_{ij}\}, \{a_i, u_i\}, \{c_{ij}, z_{u_iu_j}\}$ remain.  Consider
        vertex $b_i$, where $i \in [k]$.  Then $\{b_i\} \in \W$.  In either
        graph, $\{b_i\}$ is adjacent to $\{a_i, u_i\}$ and no other remaining
        witness set. 

    \item Consider vertex $d_{ij}$, with distinct $i, j \in [k]$.  In either
        graph, $\{d_{ij}\}$ is only adjacent to $\{c_{ij}, z_{u_i u_j}\}$ among
        the witness sets that remain to consider.

    \item The remaining witness sets have the form $\{a_i, u_i\}$ or $\{c_{ij},
        z_{u_i u_j}\}$.  Consider $i \in [k]$ and witness set $\{a_i, u_i\}$.
        Among the witness sets that remain to consider, in $H$, it is adjacent
        to all witness sets in  $\{ \{c_{ij} z_{u_iu_j}\} : j \in [k] \setminus
        \{i\}\}$, but not to any other $\{a_j u_j\}$ set nor to any other
        $\{c_{jh} z_e\}$ witness set with $j, h \neq i$.  The same is true in $G$, because $u_i$
        has every $z_{u_i u_j}, j \in [k] \setminus \{i\}$ in its neighborhood. 

    \item Finally, for distinct $i, j \in [k]$, we must consider $\{c_{ij},
        z_{u_i u_j}\}$.  In either graph, all their adjacencies were handled
        previously, as these are not adjacent to any other $\{c_{xy}, z_{u_x
        u_y}\}$ set.
    
\end{itemize}
We have thus shown that if we partition $V(G)$ into $\W$ and apply the
corresponding contractions, and do the same on $H$, we obtain the same graph.

In the converse direction, 
let $M$ be a common contraction of $G$ and $H$ achievable in a total of $K$
contractions.
We need two witness structures, one for $G$ into $M$ and one for $H$ into $M$.
So let $\X$ be a witness structure of $G$ into $M$, and $\Y$ a witness
structure of $H$ into $M$. Note that for a vertex $w$, we write $\X(w)$ for the
witness set of $\X$ that contains $w$, and $\Y(w)$ for the witness set of $\Y$
that contains $w$.  

We must consider multiple ways that $G$ and $H$ could be turned into $M$, so we
split the proof into claims.

\begin{claim}\label{claim:one-of-abcd}
    For each $i \in [k]$, at least one of $\X(a_i)$ or $\X(b_i)$ has two
    elements or more.  Likewise, for each distinct $i, j \in [k]$, at least one
    of $\X(c_{ij})$ or $\X(d_{ij})$ has two elements or more.
\end{claim}

\begin{proof}
    Suppose for contradiction that $\X(a_i) = \{a_i\}$ and $\X(b_i) = \{b_i\}$.
    Then $a_i, b_i$ are representatives in $\X$ and thus vertices of $M$.
    Since $a_i b_i \notin E(G)$, they do not share an edge in $M$.  Now
    considering $H$, we have $\Y(a_i) \neq \Y(b_i)$ since $a_i, b_i$ are both
    in $M$.  However, $\Y(a_i), \Y(b_i)$ are adjacent in $H$, and thus
    according to $\Y$ we should have the edge $a_i b_i$ in $M$, a
    contradiction.  
	
    The argument for $\X(c_{ij}), \X(d_{ij})$ is identical. Assume that
    $\X(c_{ij}) = \{c_{ij}\}$ and $\X(d_{ij}) = \{d_{ij}\}$.  Then $c_{ij},
    d_{ij}$ are in $M$ and do not share an edge since $c_{ij} d_{ij} \notin
    E(G)$.  However, $c_{ij} d_{ij} \in E(H)$, leading to the same
    contradiction as above.
\end{proof}

Claim~\ref{claim:one-of-abcd} lets us construct a matching $R \subseteq E(G)$
as follows.  For each $i \in [k]$, by Claim~\ref{claim:one-of-abcd} there is
either an edge $a_i u$ or $b_i u$ whose ends are in the same set of $\X$, where
$u$ is a neighbor of $a_i$ or $b_i$ in $G$ and thus $u \in V'_i$.  Choose any
such edge and add it to $R$ (make one choice per $i \in [k]$, which clearly
gives a matching so far).  Likewise for distinct $i, j \in [k]$, there is
either an edge $c_{ij} z$ or $d_{ij} z$ whose ends are in the same set of $\X$,
where $z \in Z'_{ij}$.  Add any such edge to $R$, for each $i, j$, and notice
that this does produce a matching $R$ such that the ends of each edge are in
the same witness set.  

By Lemma~\ref{lem:matching}, $R$ corresponds to a set of contractions we can
perform first in a sequence from $G$ to $M$.  We have $|R| = k + {k \choose
2} = K/2$, and applying the contractions removes $K/2$ vertices from $G$.  Since $V(G)
= V(H)$, at least $K/2$ vertices must also disappear from $H$ to $M$, and it
follows that $H$ also requires at least $K/2$ contractions.   Because $M$ can be
reached with a total of $K$ contractions, we deduce from
Claim~\ref{claim:one-of-abcd} that after applying the contractions
corresponding to $R$, there is no room for further contractions.  It follows
that \emph{exactly} one of $a_i$ or $b_i$ has a partner in its witness set of $\X$, and
the same holds for $c_{ij}$ versus $d_{ij}$ (we note that $a_i, b_i$ cannot be
in the same witness set of $\X$ since they are not neighbors, same for $c_{ij},
d_{ij}$).  Furthermore, the edges contracted from $G$ to $M$ must consist of
precisely $R$, although we do not know which end of each edge is kept or
removed.  However we can argue that it is the $a_i$'s and the $c_{ij}$'s that
are part of the matching, not the $b_i$'s or $d_{ij}$'s.

\begin{claim}\label{claim:only-ai-cij}
For each $i \in [k]$, $\X(a_i) = \{a_i, u\}$ for some $u \in V'_i$.  Likewise
for each distinct $i, j \in [k]$, $\X(c_{ij}) = \{c_{ij}, z\}$ for some $z
\in Z'_{ij}$.
\end{claim}

\begin{proof}
Let $i \in [k]$ and note that by Claim~\ref{claim:one-of-abcd} and the above
discussion that either $|\X(a_i)| = 2$ or $|\X(b_i)| = 2$, but not both.
Suppose for contradiction that $\X(b_i) = \{b_i, u\}$ for some $u \in
V'_i$.  Then $\X(a_i) = \{a_i\}$.  Since at most one vertex of $V'_i$ can
be part of a contraction, in $M$ every vertex of $V'_i \setminus \{u\}$ is
present.  So by inspecting $N_G(a_i)$, we infer that in $M$, the vertex $a_i$ is adjacent to $V'_i \setminus
\{u\}$, and to $u$ or $b_i$, whichever is the representative of $\X(b_i)$,
and $a_i$ has no other neighbor in $M$.

Consider vertex $c_{ij}$ of $H$ and $\Y(c_{ij})$.  We cannot have $\Y(a_i) =
\Y(c_{ij})$, as this would make $\Y(a_i)$ adjacent to all extra vertices of
$Z'_{ij}$ that are in $H$ (and such vertices exist in $M$ since there are $4k^3 >
K$ of them).  The established adjacencies of $a_i$ in $M$ do not allow
this.  Thus $\Y(c_{ij}) \neq \Y(a_i)$ and the two witness sets are adjacent
in $H$ because $a_i c_{ij} \in E(H)$.  The representative $x$ of
$\Y(c_{ij})$ must be a neighbor of $a_i$ in $M$, and so $x$ is in $V'_i
\cup \{b_i\}$.  However, this makes $x$ adjacent to all extra vertices of
$Z'_{ij}$ that remain in $M$, which is clearly not possible by applying the
contractions on $G$ of the matching $R$ discussed above.  Therefore,
$|\X(a_i)| = 2$.

Let us move on to the second part, that each $\X(c_{ij})$ has two elements.
The proof goes along the same lines.  Suppose instead that $\X(d_{ij}) =
\{d_{ij}, z\}$ for some $z \in Z'_{ij}$ and $\X(c_{ij}) = \{c_{ij}\}$.  From $N_G(c_{ij})$ we infer that in
$M$, $c_{ij}$ has only elements of $Z'_{ij} \cup \{d_{ij}\}$ in its
neighborhood.  Now consider $\Y(a_i)$ and note that $\Y(a_i) \neq
\Y(c_{ij})$ as this would make $c_{ij}$ adjacent to extra vertices of
$V'_i$.  The two witness sets are adjacent, and so the representative $x$
of $\Y(a_i)$ is in $Z'_{ij} \cup \{d_{ij}\}$.  This makes $x$ adjacent to
extra vertices of $V'_i$, which is not possible from the contractions of
$R$ on $G$.  Thus $|\X(c_{ij})| = 2$.
\end{proof}

By Claim~\ref{claim:only-ai-cij}, each $a_i$ is contracted with exactly one
$u_i \in V'_i$.  We claim that if we take all $u_i$'s in the same witness set
as an $a_i$, we get a multicolored clique of $G_C$.

\begin{claim}
    Consider distinct $i, j \in [k]$ and $u_i, u_j$ be such that $\X(a_i) =
    \{a_i, u_i\}$ and $\X(a_j) = \{a_j, u_j\}$.  Then $u_i \in V_i$, $u_j \in
    V_j$, and $u_i u_j \in E(G_C)$.
\end{claim}

\begin{proof}
    We know that $u_i \in V'_i$ and $u_j \in V'_j$.  Let us assume that either $u_i$ or $u_j$ is an extra vertex, or that $u_i
    u_j \notin E(G_C)$.  Either way, in $G$ and $H$ the vertex $z_{u_i u_j}$
    does not exist.  Now let $z \in Z'_{ij}$ such that $\X(c_{ij}) = \{c_{ij},
    z\}$, which exists by Claim~\ref{claim:only-ai-cij}.  In $G$ and $H$, at least
    one of $u_i$ or $u_j$ is not adjacent to $z$, as otherwise $z$ would be
    $z_{u_i u_j}$.  Suppose without loss of generality that $u_i$ is not
    adjacent to $z$, and
    let $x \in \{a_i, u_i\}$ be the representative of $\X(a_i)$, and $y \in
    \{c_{ij}, z\}$ the representative of $\X(c_{ij})$.  
    Then in $G$, the witness sets $\X(x) = \{a_i, u_i\}$ and
    $\X(y) = \{c_{ij}, z\}$ are not adjacent.

    Observe that in $M$,
    the neighborhood of $x$ includes all extra vertices $V'_i \setminus \{x\}$.
    Likewise, $y$ has among its neighbors in $M$ all extra vertices of $Z'_{ij}
    \setminus \{y\}$.
    One implication is that in $H$, $\Y(x)$ must be adjacent to all witness
    sets of $\Y$ that contain an extra vertex of $V'_i$ (except possibly
    $\Y(x)$ itself).  Since there are $4k^3 > 2K$ extra vertices, this is only
    possible if $\Y(x)$ has a vertex that is a neighbor of all extra vertices,
    which is either $b_i$ or $a_i$.  Since $b_i$ is in $M$, $b_i \notin \Y(x)$,
    and thus $a_i$ is in $\Y(x)$.  Using the same reasoning on $y$ and
    $c_{ij}$, we get that $c_{ij}$ is in $\Y(y)$.  Hence $\Y(x)$ and $\Y(y)$
    are adjacent in $H$, a contradiction since $\X(x)$ and $\X(y)$ are not
    adjacent in $G$.
\end{proof}

By this last claim, we can take the set $\{ u_i : u_i \in \X(a_i), i \in [k]
\}$, and all these vertices are in $G_C$ and every pair shares an edge.  Thus
$G_C$ has a multicolored clique.
\end{proof}
\end{toappendix}

\subsection{Parameterization by $k + \Delta$}

Let $G$ be a graph of maximum degree $\Delta$.  It is not hard to see that
applying at most $k$ contractions to $G$ results in a graph of maximum degree
$O(k\Delta)$, which we make precise in the next lemma.
It then becomes relatively easy to get an FPT algorithm in $k + \Delta$.

\begin{lemma_rep}\label{lem:maxdeg}
Let $G$ be a graph of maximum degree $\Delta \geq 2$.  Then applying $t
\geq 0$ contractions to $G$ results in a graph of maximum degree at most
${\Delta + t (\Delta - 2)}$.

Moreover, this bound is tight, that is, for any $t$, there exists a graph
$G$ of maximum degree $\Delta$ and a sequence of $t$ contractions that
results in a graph of maximum degree $\Delta + t(\Delta - 2)$.
\end{lemma_rep}


\begin{proof}
For the first part, let $G$ have maximum degree at most $\Delta$ and let $H$ be
a graph obtained after $t$ contractions, with $\W$ the witness structure of
$G$ into $H$.  Let $W \in \W$ be one of the witness sets with
representative $v \in V(H)$ (so that $W_v = \W(v)$).  In $H$, the degree of
$v$ is at most the number of edges of $G$ with one endpoint in $W_v$ and
the other outside $W_v$.  Let $e_{out}$ denote this number of edges, and
let $e_{in} = |E(G[W_v])|$.  Since $G[W_v]$ is connected, $e_{in} \geq
|W_v| - 1$.  In $G$, the sum of degrees of the $W_v$ vertices is equal to
$e_{out} + 2e_{in}$, and this quantity is at most $\Delta|W_v|$.
Rearranging a bit, \[ e_{out} \leq \Delta|W_v| - 2e_{in} \leq \Delta|W_v| -
2(|W_v| - 1) = |W_v|(\Delta - 2) + 2.  \]  Since $\Delta \geq 2$ and $|W_v|
\leq t + 1$, replacing $|W_v|$ with $t + 1$ in the previous expression
yields the desired upper bound on the degree of an arbitrary vertex $v$ of
$H$.

As for tightness, construct a graph $G$ as follows.  Start with a path with $t
+ 1$ vertices, then for each vertex $v$ add $\Delta - 2$ new vertices of
degree $1$ whose neighbor is $v$.  In addition, we add an extra degree $1$
neighbor to each end of the path.  One can see that contracting the initial
path into a single vertex applies $t$ contractions, and results in a star
whose center has degree $(t + 1) (\Delta - 2) + 2 =  \Delta + t(\Delta -
2)$.
\end{proof}

\begin{proposition_rep}
The \textsc{Maximum Common Labeled Contraction} problem can be solved in time
    $O((8(k+1)\Delta)^{k+1} \cdot n)$, where $k$ is the number of contractions,
    $\Delta = \max(\Delta(G), \Delta(H))$, ${n = |V(G)| + |V(H)|}$ and $m = |E(G)| + |E(H)|$.
\end{proposition_rep}

\begin{proofsketch}
    Consider a recursive branching algorithm as follows.  If there exists a vertex $u$ present in one graph but no the other, we know that $u$ must be incident to a contraction and must disappear.   By Observation~\ref{obs:witness-order}, we may do this contraction first.  By Lemma~\ref{lem:maxdeg}, $u$ has degree at most $\Delta + k(\Delta - 2)$, which we bound by $(k + 1)\Delta$ for simplicity.  We branch on all the ways of contracting $u$ with a neighbor.  
    If no such $u$ exists, then $V(G) = V(H)$.  If $E(G) = E(H)$, we are done, so assume there is $uv \in E(G) \setminus E(H)$.  We observe that $u$ or $v$ must be incident to some contraction, although it could be in either graph, and $u/v$ could be the kept vertex or not.  There are $8 (k + 1) \Delta$ ways to branch on such contractions. 
 Since each branch reduces the parameter $k$ by $1$, we get a recursive search tree with $(8(k+1)\Delta)^k$ calls.  We add a factor proportional to $k \Delta n$ to check all the edges at each recursion.
\end{proofsketch}

\begin{proof}
Consider an instance of \textsc{Maximum Common Labeled Contraction} consisting
of labeled graphs $G, H$ of maximum degree $\Delta$ or less, and parameter
$k$ for the allowed number of contractions.  We describe a recursive
branching algorithm in which each branching path makes at most $k$
contractions (or otherwise returns that no solution is possible in this
path).  By Lemma~\ref{lem:maxdeg}, any instance encountered in a recursion
has maximum degree at most $\Delta + k(\Delta - 2)$, which we upper bound
by $(k+1)\Delta$ for simplicity. 

So we assume that we are in a recursive call which receives $G$ and $H$ and
integer $k$.  Suppose that there is some $v \in V(G)$ such that $v \notin
V(H)$.  Then to obtain any common contraction, $v$ must be in the same
witness set as one of its neighbors.  We branch on every possible
contraction $(w, v)$, for every neighbor $w$ of $v$.  We decrease the
parameter $k$ by $1$ in each branch.  This branches into at most $(k+1)\Delta$
cases, and by Observation~\ref{obs:witness-order} one of these branches
will lead to a maximum common contraction with $k$ operations if one
exists.  The same branching can be applied if $H$ has some $v$ not in $G$.  

So assume that $V(G) = V(H)$.  If $E(G) = E(H)$ then we are done, so assume
without loss of generality that $G$ has an edge $uv$ that is not in $H$.
Observe that to obtain a common contraction, $u$ or $v$ must be incident to
at least one contraction, in at least one of $G$ or $H$, as otherwise the
edge $uv$ stays present in $G$ and it remains a non-edge in $H$.  That is,
one of $u$ or $v$ has a neighbor in the same witness set, in one of the
graphs.  By Observation~\ref{obs:witness-order}, we can  branch into all
the contractions that $u$ or $v$ could be part of, in either graph, i.e.,
we branch into every contraction $(x, y)$ where $\{u, v\} \cap \{x, y\}
\neq \emptyset$, and where $xy \in E(G) \cup E(H)$.   The number of such
possible contractions incident to $u$ is at most $2(|N_G(u)| + |N_H(u)|)
\leq 2((k+1)\Delta + (k+1)\Delta) = 4(k+1)\Delta$ (we multiply by $2$ because
contractions are ordered pairs).  The same bound holds for $v$, so in total
we branch into at most $8(k+1)\Delta$ possibilities, still decreasing $k$ by
$1$ in each case.

It is easy to see that the algorithm finds a common contraction achievable with
at most $k$ contractions if and only if one exists.  For the complexity,
the algorithm creates a recursion tree in which each vertex has at most
$8(k+1)\Delta$ children, and whose depth is at most $k$.  In each recursive
call, there are at most $2k\Delta \cdot n$ edges, so we can check in time
$O(k\Delta n)$ whether $G$ and $H$ are the same graph, and if not find a
vertex or a vertex pair to branch on.  This results in a running time of
$O((8(k+1)\Delta)^{k+1} \cdot n)$.
\end{proof}



\subsection{Tractability when parameterizing by the treewidth of $G \cup H$}

We now prove that \textsc{Maximum Common Contraction} is FPT in
$tw(G\cup H)$, i.e.\ the treewidth of the graph with vertices
 $V(G) \cup V(H)$, and edges $E(G)\cup E(H)$.
We remark that $tw(G \cup H)$ could be unbounded even if both $G, H$
have bounded treewidth~\cite{alecu2024treewidth}, but nonetheless we derive
consequences from our result at the end of the section.
We use
Courcelle's theorem~\cite{courcelle1990monadic} by formulating the
problem in the monadic second-order logic on graphs. More precisely, we
use an extended version of Courcelle's theorem~\cite{arnborg1991easy}
allowing for \emph{labeled edges} and \emph{labeled vertices}
if labels are taken from a finite set.
Indeed, in $G\cup H$, we still need to distinguish vertices in $
V(G)$ and $V(H)$,
and edges in $E(G)$ and $E(H)$.
Alternatively, following the terminology of \cite{flum2006parameterized},
we are applying Courcelle's theorem on a \emph{relational structure} 
(Theorem 11.37 in \cite{flum2006parameterized}) on $V(G)\cup V(H)$,
containing two unary relations labeling $V(G)$ and $V(H)$, and two
binary relations for $E(G)$ and $E(H)$. One can verify that the
treewith of this relational structure (in the sense of \cite[Definition
11.23]{flum2006parameterized}) is $tw(G\cup H)$.

\begin{theorem}
\textsc{Maximum Common Conraction} on fully-labeled graphs is fixed-parameter tractable 
in $tw(G\cup H)$.
\label{thm_g1g2twcourcelle}
\end{theorem}
\begin{proof}
    In this proof, we denote $V(G)$ by $V_1$, $V(H)$ by $V_2$, $E(G)$ by
    $E_1$ and $E(H)$ by $E_2$ for simplicity.  We use the following
    formulation of the problem: we are looking for sub-sets $S_1\subseteq E_1$
    and $S_2\subseteq E_2$ of edges, and a subset $R\subseteq V_1\cap V_2$ of
    vertices (``representatives'') such that $S_1$ and $S_2$ induce, in $G$
    and $H$ respectively, connected components (the witness sets) and such that
    each connected component contains exactly one vertex in $R$. 

    This can be expressed in MSOL in the following way: \begin{align*} \exists
        S_1\subseteq E_1, &\exists S_2\subseteq E_2, \exists R\subseteq V_1\cap
        V_2\text{ s.t }\\ &\textbf{contraction}_1(S_1,R)\text{ and
        }\textbf{contraction}_2(S_2,R)\text{ and }\\ &\forall r,r'\in R \textbf{
    edge}_1 (S_1,r,r')\text{ iff }\textbf{edge}_2(S_2,r,r') \end{align*} Where
in this formula, for $i=1,2$ $\textbf{contraction}_i(S_i,R)$ checks that that
    there is exactly one element of $R$ per connected component of $V_i$
    induced by $S_i$, and $\textbf{edge}_i(S_i,r,r')$ is true if and only there
    is an edge in $E_i$ between the connected components induced by $S_i$ on
    $V_i$ containing $r$ and $r'$, respectively. We express
    $\textbf{contraction}_i(S_i,R)$ as follows: \begin{align*}
        \textbf{contraction}_i(S_i,R) &= \forall r,r'\in R, r\neq r',
        \neg\textbf{path}(r,r',S_i) \text{ and }\forall u\in V_i\setminus
        R\;\exists r\in R\text{ s.t. }\\ & \textbf{path}(u,r,S_i)\text{ and
    }\forall r'\neq r\;\neg\textbf{path}(u,r',S_i) \end{align*} Where in this
    formula, \textbf{path}$(x,y,U)$ checks whether $x=y$ or there exists a path between
    $x$ and $y$ using only edges in $U$ (which can be checked in
    MSOL~\cite{courcelle1990graph}) if $x\neq y$. Indeed, two vertices are in the same
    connected component induced by a sub-set of edges if and only there is a
    path between them using only edges from the sub-set.

    As for $\textbf{edge}_i(S_i,r,r')$, it can be expressed as:
    \begin{align*}
        \textbf{edge}_i(S_i,r,r') &= \exists x,y\in V_i\text{ s.t. } \textbf{path}(x,r,S_i)\text{ and }\textbf{path}(y,r,S_i)\text{ and } \{x,y\}\in E_i \end{align*}
    Overall, this expression only uses quantification over edge subsets
        ($S_1,S_2$) and vertex sub-sets ($R$). By the extended version of
        Courcelle's theorem allowing to optimize for the size of monadic
        variables (\cite{arnborg1991easy} or \cite[Exercise 11.44]{flum2006parameterized}), one can find
        $S_1,S_2,R$ verifying this formula while maximizing $|R|$ (or
        equivalently minimizing $|S_1|+|S_2|$) in a complexity FPT in the treewidth of $(V_1\cup V_2,E_1\cup E_2)=G\cup H$.
\end{proof}

We now derive consequences of Theorem~\ref{thm_g1g2twcourcelle} by establishing parameterizations that lead $G \cup H$ to have bounded treewidth.
In the following, we have graphs $G$ and $H$ and $G\cup H$ is the graph with vertex set $V(G)\cup V(H)$
and edge set $E(G)\cup E(H)$. 
For space reasons, the definition of a tree decomposition can be found in the appendix.

\begin{toappendix}
    Let $G$ be a graph.  A \emph{tree decomposition} of $G$ is a pair $\T=(T,\{X_t\}_{t\in V(T)})$ where $T$ is a tree and, for each $t \in V(T)$, $X_t \subseteq V(G)$.  The sets $X_t$ are called \emph{bags}.  Moreover, $\T$ satisfies the following:
    \begin{enumerate}
        \item 
        for each $x \in V(G)$, some bag $X_t$ of $\T$ contains $x$.

        \item 
        for each $x \in V(G)$, the vertices of $T$ corresponding to bags that contain $x$ form a connected subtree of $T$.

        \item 
        for each edge $uv \in E(G)$, some bag contains both $u$ and $v$.
    \end{enumerate}
    The \emph{width} of $\T$ is $\max_{t \in V(T)} (|X_t| - 1)$, and the \emph{treewidth} of $G$ is the minimum width of a tree decomposition of $G$.
\end{toappendix}

\begin{lemma_rep}
    If $H$ is a contraction of $G$, then $tw(G\cup H)\leq 2 \cdot tw(G)$.
\end{lemma_rep}

\begin{proofsketch}
    Start with a tree decomposition $\T=(T,\{X_t\}_{t\in V(T)})$ of $G$.  Then take a witness structure $\W$ of $G$ into $T$.  For each bag $X_t$ of $\T$, and for each $x \in X_t$, add to $X_t$ the representative of $\W(x)$ in $H$, if not already there.  This at most doubles the treewidth.  The bags that contain some $u \in V(G) \cup V(H)$ are connected in the modified decomposition, as otherwise a bag that separates $u$ would also separate $\W(u)$, which should be connected.  Also, for each $uv \in E(H)$ some bag has both $u, v$ because $\W(u), \W(v)$ are adjacent.
\end{proofsketch}

\begin{proof}
    Let $\T=(T,\{X_t\}_{t\in V(T)})$
    be a tree decomposition of $G$, and $\W$ a witness structure of $G$ into $H$.
    We build a new decomposition $\T'=(T,\{Z_t\}_{t\in T})$
    and argue it is a tree decomposition of $G\cup H$.
    The underlying tree $T$ is the same for $\T$ and $\T'$.
    As for the bags, we set:
    $$Z_t=X_t\cup\{u\in V(H) : \exists x\in X_t\text{ s.t }x\in \W(u)\}$$

    In other words, 
    we take each bag $X_t$, and for each $x \in X_t$ we add the representative in $H$ of the witness set of $x$, which results in $Z_t$.
    Let us check that $\T'$ is a valid tree decomposition of $G\cup H$.
    \begin{itemize}
        \item (connectivity of vertex representation)
            First note that a vertex $x \in V(G) \setminus V(H)$ belongs to $Z_t$ if and only if $x$ belongs to $X_t$, since only vertices of $H$ are added to bags.  Since the bags containing $x$ are connected in $\T$, they are still connected in $\T'$.
            Now 
            Let $u\in V(H)$ and $t\in V(T)$. 
            First, note that since $u\in V(G)$, there is at least
            one bag of $\T$ and thus $\T'$ containing it. Then,
            let $t_1,t_2,t_3$ be three vertices of $T$ such that $t_3$ is on the path from $t_1$
            to $t_2$. Suppose that $u\in Z_{t_1}$, $u\in Z_{t_2}$ but $u\notin Z_{t_3}$.
            By definition of $\T'$, $\exists a\in X_{t_1}$ such that $a\in \W(u)$
            and $b\in X_{t_2}$ such that $b\in \W(u)$.
            Note that $a = u$ is possible, in which case $u \in X_{t_1}$ (likewise, $b = u$ is possible).
            We also have $\W(u)\cap X_{t_3}=\emptyset$, as otherwise $u$ would be added to $Z_{t_3}$, and 
            in particular, $a, b \notin X_{t_3}$.
            
            If $a=b$, we have a contradiction with $\T$ being a valid tree decomposition of $G$,
            since $a \in V(G)$ and the bags of $\T$ containing it would be disconnected by $t_3$. If $a\neq b$,
            since $\T$ is a tree decomposition of $G$, $X_{t_3}$ separates $a$ from $b$,
            i.e.\ intersects all paths from $a$ to $b$ in $G$. However, by the connectivity of $\W(u)$,
            there is a path from $a$ to $b$ lying entirely in $\W(u)$, which yields a contradiction.
        \item (edge representation) 
        For $uv \in E(G)$, some bag of $\T$ contains $u$ and $v$, and the same holds in $\T'$. Then,  
        for each edge $uv$ of $H$, there
            must be at least one edge $xy$ of $G$ such that $x\in \W(u)$,
            $y\in \W(v)$. There must be a bag in $\T$ containing both
            $x$ and $y$. By definition, this bag contains $u$ and $v$
            in $\T'$, therefore representing the edge.
    \end{itemize}
By definition of $\T'$, each of its bag $Z_t$ is at most twice
    as big as $X_t$, because each $x \in X_t$ enforces the addition of at most one new vertex in $Z_t$, namely the $u \in V(H)$ that is the representative of its witness set. Therefore the width of $\T'$ is at most $2 tw(G)$.
\end{proof}

\begin{lemma_rep}
    If $G$ and $H$ are such that 
    $\exists S_1,S_2$
    labeled contraction sequences of size $|S_1|+|S_2|\leq k$ 
    and $G/S_1=H/S_2$, then $tw(G\cup H)\leq \min(tw(G),tw(H))+k$
\end{lemma_rep}

\begin{proofsketch}
    Let $M$ be a maximum common contraction of $G$ and $H$.  Since contractions do not increase the treewidth, $tw(M) \leq \min(tw(G), tw(H))$.  
    Then, take a tree decomposition $\T$ of $M$, then add $(V(G) \setminus V(H)) \cup (V(H) \setminus V(G))$ to every bag.  This adds at most $k$ vertices to each bag, and one can argue that this is a valid tree decomposition of $G \cup H$.
\end{proofsketch}

\begin{proof}
    Let $S_1,S_2$ be as in the statement and $M=G/S_1=H/S_2$. Let
     $\T=(T,\{X_t\}_{t\in V(T)})$ be a tree decomposition of $M$.
    We modify $\T$ into $\T'=(T,\{Z_t\}_{t\in V(T)})$, a tree decomposition
    with the same underlying tree $T$, and $\forall t\in V(T)$, $Z_t=X_t\cup \left((V(G)\cup V(H))\setminus V(M)\right)$. In other words, we add all deleted vertices to all bags of $\T$.
    Since $|S_1|+|S_2|=|(V(G)\cup V(H))\setminus V(M)|\leq k$, we have that $\forall t\in V(T)\,|Z_t|\leq|X_t|+k$. It remains to argue that $\T'$ is a tree decomposition of $G\cup H$, and conclude since
    $tw(M)\leq \min(tw(G),tw(H))$ (from being a contraction of both graphs).
    \begin{itemize}
        \item Let $x\in V(G)\cup V(H)$. If $x\in V(M)$, then the set of bags containing $x$
        in $\T'$ is the same as in $\T$, i.e.\ non-empty and connected by the validity of $\T$. 
        If $x\notin V(M)$, then all bags contain $x$.
        \item Let $x,y$ be an edge of $G$ (the argument is the same for an edge of $H$).
        If both $x,y$ are in $V(M)$, then $x,y$ is an edge of $M$, and by the validity of $\T$,
        there is $t\in V(T)$ such that $\{x,y\}\subseteq X_t\subseteq Z_t$. If $x\notin V(M)$ 
        then it is present everywhere, and in particular in bags containing $y$. The same argument
        applies if $y\notin M$.
    \end{itemize}
\end{proof}

\begin{corollary}
    The following results holds:
    \begin{itemize}
        \item 
        \textsc{Labeled Contractibility} is FPT in parameter $\max(tw(G), tw(H))$.
        \item 
        \textsc{Maximum Common Labeled Contraction} is FPT in parameter $k + \min(tw(G), tw(H))$. 
    \end{itemize}
\end{corollary}

\medskip 

\noindent
\textbf{Conclusion.} 
We have explored the parameterized complexity of computing
a maximum common contraction of two fully-labeld graphs,
in which each label is used at most once in each graph.
Natural follow-ups could look at the case where labels
can be used more than once. Note that it is the hardest possible
case, as it contains both unlabelled and uniquely labeled graphs.
A first step could be to bound the number of times each label may be 
present. As for the unlabeled case, it consists in computing
edge-contraction distances between graph isomorphism classes,
and has not been explored yet.
We finish with some open problems: (1) given
Theorem~\ref{thm_cw}, is
\contractibilitypb still NP-hard if both input graphs have clique-width three,
or even if they are both cographs (clique-width two)? (2) is there a
$2^{O(\delta k)} n^c$ time algorithm for \contractibilitypb, and/or a
$2^{O(\Delta k)} n^c$ time algorithm for \maxcommonpb?  Or better, could there
be subexponential algorithms parameterized by $\delta k$ or $\Delta k$?  (3) is
\maxcommonpb FPT in $\max(tw(G), tw(H)$?

\newpage

\bibliographystyle{plainurl}
\bibliography{biblio}

\appendix

\end{document}